\documentclass[prb,twocolumn,10pt,aps,showpacs,floatfix,amsmath,amsfonts,letterpaper,longbibliography,nofootinbib]{revtex4-2}

\usepackage{stix} 
\usepackage{graphicx}
\usepackage{amsmath}
\usepackage{url}
\usepackage{array}
\usepackage{multirow}
\usepackage{bm}

\newcolumntype{C}{>{\centering\arraybackslash}p{2.5cm}}
\newcolumntype{P}[1]{>{\centering\arraybackslash}p{#1}}

\makeatletter
\def\beq{\@ifstar{\@ifnextchar[{\@beqslabel}{\@beqsnolabel}}
{\@ifnextchar[{\@beqlabel}{\@beqnolabel}}}
\def\@beqlabel[#1]{\begin{equation}\label{#1}}
\def\@beqnolabel{\begin{equation}}
\def\@beqslabel[#1]{\begin{equation*}\label{#1}}
\def\@beqsnolabel{\begin{equation*}}
\def\eeq{\@ifstar{\end{equation*}}{\end{equation}}}
\makeatother

\newcommand{\refcite}[1]{Ref.~\cite{#1}}
\newcommand{\refcites}[1]{Refs.~\cite{#1}}

\newcommand{\refeq}[1]{Eq.~(\ref{#1})}

\newcommand{\refeqand}[2]{Eqs.~(\ref{#1}) and (\ref{#2})}

\newcommand{\reffig}[1]{Fig.~\ref{#1}}

\newcommand{\refsec}[1]{Sec.~\ref{#1}}
\newcommand{\refapp}[1]{Appendix~\ref{#1}}

\newcommand{\reftab}[1]{Table~\ref{#1}}

\newcommand{\hc}[1]{{#1}^{\dagger}}
\newcommand{\dx}{\partial_{x}}
\newcommand{\ee}{\mathrm{e}}
\newcommand{\ii}{\mathrm{i}}
\newcommand{\dd}{\mathrm{d}}
\newcommand{\del}{\bm{\nabla}}

\newcommand{\punc}[1]{\,{\text{#1}}}
\newcommand{\sub}[1]{_{\text{#1}}}
\newcommand{\zero}{^{(0)}}
\newcommand{\spr}[1]{^{(#1)}}

\newcommand{\rv}{\bm{r}}
\newcommand{\Rv}{\bm{R}}
\newcommand{\deltav}{\bm{\delta}}
\newcommand{\goC}{\mathfrak{C}}

\newcommand{\zerov}{\bm{0}}
\newcommand{\Phiv}{\bm{\Phi}}

\newcommand{\ham}{\mathcal{H}}
\newcommand{\vev}[1]{\langle  {#1} \rangle_{0}}

\DeclareMathOperator{\Tr}{Tr}

\newcommand{\no}[1]{\mathopen{:}{#1}\mathclose{:}}
\newcommand{\bno}[1]{\mathopen{{}_*^*}{#1}\mathclose{{}_*^*}}

\newcommand{\Phii}{\chi} 
\newcommand{\thetai}{\vartheta} 
\newcommand{\phiz}{\phi\zero} 
\newcommand{\anis}{w} 

\usepackage[T2A,T1]{fontenc}
\newcommand{\Sha}{\mbox{\usefont{T2A}{\rmdefault}{m}{n}\CYRSH}}

\usepackage[usenames,dvipsnames]{color}

\begin{document}

\title{Derivation of field theory for the classical dimer model using bosonization}

\author{Neil Wilkins}
\affiliation{School of Physics and Astronomy, The University of Nottingham, Nottingham, NG7 2RD, United Kingdom}

\author{Stephen Powell}
\affiliation{School of Physics and Astronomy, The University of Nottingham, Nottingham, NG7 2RD, United Kingdom}

\begin{abstract}
We derive a field theory for the two-dimensional classical dimer model by applying bosonization to Lieb's (fermionic) transfer-matrix solution. Our constructive approach gives results that are consistent with the well-known height theory, previously justified based on symmetry considerations, but also fixes coefficients appearing in the effective theory and the relationship between microscopic observables and operators in the field theory. In addition, we show how interactions can be included in the field theory perturbatively, treating the case of the double dimer model with interactions within and between the two replicas. Using a renormalization-group analysis, we determine the shape of the phase boundary near the noninteracting point, in agreement with results of Monte Carlo simulations.
\end{abstract}

\maketitle

\section{Introduction}

In statistical mechanics, certain lattice models with hard constraints exhibit so-called ``Coulomb phases'' \cite{Henley2010}, unconventional disordered phases with algebraic correlations, topological order, and fractionalized defects. These include vertex, coloring, and spin models \cite{Anderson1956,Youngblood1981,Blote1982,Zeng1997,Kondev1996,Castelnovo2012,Chalker2017}, as well as dimer models on bipartite lattices \cite{Huse2003,Alet2005,Alet2006}. In a long-wavelength description, the microscopic constraint becomes a continuum Gauss law, in terms of which defects are effective charges.

In three spatial dimensions and above, the constraint can be resolved in terms of an effective gauge field \cite{Huse2003}, whereas in two dimensions (2D), the corresponding resolution is in terms of a scalar ``height'' \cite{Blote1982,Henley1997}. Configurations in the 2D microscopic models can be put in (many-to-one) correspondence with an appropriately defined discrete-valued height on the dual lattice, which encodes the hard constraints in a way amenable to coarse-graining. Based on its nontrivial transformation properties under the symmetries \cite{Alet2006b}, one can write down a field theory in terms of the coarse-grained height.

The simplest massless Gaussian action for the gauge or height field then gives a continuum theory with algebraic correlations and charges subject to Coulomb-law interactions at long distances. (It has indeed been proved that such a theory describes the 2D dimer model \cite{Kenyon2001,deTiliere2007}, even in the presence of certain interactions \cite{Giuliani2015,Giuliani2017}.) Phase transitions from this Coulomb phase can be described by adding terms to the action, which, in 2D, take the form of sinusoidal functions of the height. If these ``cosine terms'' are relevant under the renormalization group (RG) \cite{Cardy1996} at the Gaussian fixed point, then the result is a Berezinskii--Kosterlitz--Thouless (BKT) transition \cite{Berezinskii1971,Kosterlitz1973,Kosterlitz2016} to a phase where fluctuations of the height are suppressed. This corresponds to an ``ordered'' phase of the microscopic model \cite{Alet2005,Alet2006b,Wilkins2020,Desai2021}, where correlations are short-ranged and defects are confined, while symmetry may or may not be spontaneously broken.

A feature special to 2D is that many of the relevant lattice models, including dimer models on planar graphs and with periodic boundaries \cite{Kasteleyn1961,Temperley1961,Fisher1961}, are exactly solvable. In particular, using Pfaffian methods and a related transfer-matrix solution \cite{Lieb1967}, it is possible to calculate exact results for dimer--dimer correlation functions \cite{Fisher1963}. These results have previously been used to fix the coefficients in the effective height-field theory \cite{Fradkin2004,Papanikolaou2007,Tang2011,Fradkin2013}, by requiring that it reproduces the asymptotic long-distance forms of the exact results.

In this work, we use the exact solution to provide a direct and explicit route to the continuum theory. Starting from the transfer-matrix solution of the square-lattice dimer model in terms of free fermions on a lattice in $1+1$ dimensions \cite{Lieb1967,Grande2011,Wilkins2021}, we constructively derive the bosonic height-field theory, including the effective long-wavelength action and the correspondence between microscopic dimer observables and operators in the field theory. As previously noted \cite{Papanikolaou2007,Giuliani2017b}, the relationship between the fermionic and bosonic theories is an instance of the general phenomenon of bosonization \cite{vonDelft1998}, and we show how this can be made concrete.

Besides its intrinsic appeal, our constructive approach has a number of benefits. By providing a direct calculation of the coefficients appearing in the field theory, it shows how they change with modifications to the microscopic model, such as including weights for dimers on horizontal and vertical links. In addition, it clarifies the origin of the short-distance cutoff required to regularize the field theory and the dependence of the bare parameters on this cutoff, which must be included correctly in order to calculate asymptotic dimer--dimer correlations.

The bosonization approach also allows certain types of interactions to be treated using perturbation theory. Since it provides precise values for coefficients in the field theory, rather than requiring them to be fixed by comparison with exact results, it leads to quantitative predictions in terms of the microscopic perturbations in the original dimer model. As an example, we consider the double dimer model \cite{Wilkins2020} and show how interactions within and between the two replicas generate cosine terms in the field theory. By treating these terms using a renormalization group (RG) analysis, we predict the shape of the phase boundary in the vicinity of the noninteracting point, in quantitative agreement with our previous computational results.

\subsection*{Outline}

In \refsec{reviewoftransfermatrixsolution}, we summarize the relevant results of the transfer-matrix method from \refcite{Wilkins2021}. These are then used to derive the field theory in \refsec{derivationoffieldtheory}. We extend the theory to the interacting case in \refsec{interactingdoubledimermodel}, allowing us to predict the phase boundary of the interacting double dimer model in the vicinity of the non-interacting point. We conclude in \refsec{conclusions}.

\section{Review of transfer-matrix solution}
\label{reviewoftransfermatrixsolution}

In this section, we review the main steps in the solution of the classical dimer model using the transfer matrix \cite{Lieb1967}. We follow the presentation in \refcite{Wilkins2021}, focusing on those results that will be used in \refsec{derivationoffieldtheory} to derive the bosonic field theory.

\subsection{Dimer model}
\label{dimermodel}

The partition function for the classical dimer model on an \(L_x\times L_y\) square lattice with periodic boundaries is
\begin{equation}
\label{eq:partitionfunction0}
Z = \sum_{c\in\goC_0} \anis^{N_{x}} \punc ,
\end{equation}
where $\goC_0$ denotes the set of all close-packed dimer configurations. We include a weight \(\anis\) assigned to each horizontal dimer; \(\anis = 1\) is the isotropic model where all dimer configurations have equal weight in the ensemble. (The parameter \(\anis\) is called \(\alpha\) in \refcites{Lieb1967,Wilkins2021}.) For simplicity, we assume throughout that \(L_x\) and \(L_y\) are both even.

Denoting by $d_{\mu}(\rv)$ the dimer occupation number (equal to \(0\) or \(1\)) on the bond joining sites $\rv$ and $\rv + \deltav_\mu$, with \(\deltav_\mu\) a unit vector in direction $\mu \in \{x,y\}$, the flux is given by
\begin{equation}
\label{eq:fluxdef}
\Phi_{\mu} = \frac{1}{L_{\mu}}\sum_{\rv}(-1)^{r_x + r_y}d_{\mu}(\rv) \punc.
\end{equation}
As a result of the close-packing constraint \cite{Chalker2017}, this is in fact equal to a sum over any single column (for \(\Phi_x\)) or row (\(\Phi_y\)),
\begin{align}
\label{eq:xflux}
\Phi_{x} &= \sum_{r_y=1}^{L_{y}}(-1)^{r_x + r_y}d_{x}(\rv)\quad\text{(any \(r_x\))}\\
\Phi_{y} &= \sum_{r_x=1}^{L_{x}}(-1)^{r_x + r_y}d_{y}(\rv)\quad\text{(any \(r_y\)),}
\label{eq:yflux}
\end{align}
and so the components of \(\Phiv\) are integers.

\subsection{Transfer matrix}

We first define a Hilbert space whose basis states correspond to all configurations of a single row of vertical bonds. The transfer matrix \(V\) is an operator that acts on any such state and gives a linear combination of possible configurations for the next row, with appropriate weights. Representing occupied and empty bonds as spin up $\lvert \uparrow \rangle$ and down $\lvert \downarrow \rangle$ respectively, we can write \cite{Lieb1967}
\begin{equation}
V = \exp\left(\anis\sum_{j=1}^{L_{x}}\sigma_{j}^{-}\sigma_{j+1}^{-}\right)\prod_{j=1}^{L_{x}}\sigma_{j}^{x}\punc,
\end{equation}
where \(\sigma_{j}^{\mu}\) and \(\sigma_{j}^{\pm} = \frac{1}{2}(\sigma_{j}^{x} \pm i \sigma_{j}^{y})\) are Pauli operators acting on the bond with column index \(j\).

In terms of the transfer matrix, the partition function, \refeq{eq:partitionfunction0}, can be expressed as
\begin{equation}
\label{eq:partitionfunction}
Z = \Tr V^{L_y}\punc.
\end{equation}
It is convenient to define the Hamiltonian \(\ham\) in terms of the two-row transfer matrix,
\beq[eq:defineham]
V^2 = \ee^{-2\ham}\punc,
\eeq
and so
\begin{equation}
\label{eq:partitionfunction2}
Z = \Tr \ee^{-L_{y}\ham}
\end{equation}
is effectively the partition function for a quantum system at inverse temperature \(L_y\). (In \refcite{Wilkins2021}, a parameter \(\bm{t}\) is included that couples to \(\Phiv\); the transfer matrix is then \(V\) and \(V^\dagger\) on even and odd rows, respectively. Here, we set \(\bm{t} = \zerov\) and so \(V = V^\dagger\). Including \(\bm{t}\) would give edge weights as in \refcite{Giuliani2020}.)

The Hamiltonian can be expressed exactly in terms of free fermions by using a Jordan--Wigner transformation to real-space fermions \(C_j\),
\begin{align}
\label{eq:JW1}
&C_{j} = \left( \prod_{i=1}^{j-1}- \sigma_{i}^{z} \right)\sigma_{j}^{-} \\
\label{eq:JW3}
&C_{j}^{\dagger}C_{j} = \frac{1}{2}(1 + \sigma_{j}^{z}) \punc ,
\end{align}
with boundary condition
\begin{equation}
\label{eq:cbcs}
C_{L_{x} + 1} = -(-1)^{p}C_{1}\punc,
\end{equation}
where \(p = 0\) (even) or \(1\) (odd) is the parity of the total fermion number.

This is followed by a Bogoliubov transformation to momentum space fermions \(\zeta_k\),
\begin{equation}
\label{eq:czeta}
C_{j} =
\sqrt{\frac{2}{L_{x}}}\ee^{-\ii \pi/4}\sum_{k\in\mathbb{K}_{p}}\ee^{\ii kj}\cos\theta_{k} \times \begin{cases}
\zeta_{k} & \text{for \(j\) even}\\
\zeta_{-k}^{\dagger} & \text{\phantom{for} \(j\) odd,}
\end{cases}
\end{equation}
where
\begin{equation}
\label{eq:angle}
\tan(2\theta_{k}) = \frac{1}{\anis \sin k} \punc, \qquad \theta_{k} \in \left[0 ,\frac{\pi}{2}\right]
\end{equation}
and \(\mathbb{K}_0\) (\(\mathbb{K}_1\)) is the set of all half-integer (resp., integer) multiples of \(2\pi/L_x\) in \([-\pi,\pi)\), i.e.,
\begin{align}
\label{eq:qevenflux}
\mathbb{K}_{0} &= \{\pm\pi/L_{x}, \pm 3\pi/L_{x}, \dotsc , \pm(L_{x}-1)\pi/L_{x}\}\\
\label{eq:qoddflux}
\mathbb{K}_{1} &= \{0, \pm 2\pi/L_{x}, \pm 4\pi/L_{x}, \dotsc , \pm(L_{x}-2)\pi/L_{x}, -\pi\} \punc .
\end{align}

The Hamiltonian in parity sector \(p\) is then
\begin{equation}
\label{eq:ham2}
\ham_{p} = \sum_{k\in\mathbb{K}_{p}}\epsilon(k)\zeta_{k}^{\dagger}\zeta_{k} \punc ,
\end{equation}
where
\begin{equation}
\label{eq:dispersion}
\epsilon(k) = \sinh^{-1}(\anis\sin k) \punc.
\end{equation}
The full set of eigenstates of the original Hamiltonian \(\ham\) is given by the union of those eigenstates of \(\ham_0\) that have even total (real-space) fermion number and those of \(\ham_1\) that have odd total number.

The \(y\) component of the flux \(\Phiv\) can similarly be written as
\begin{equation}
\label{eq:fluxzeta}
\Phi_{y} = -\frac{L_{x}}{2} + \sum_{k\in\mathbb{K}_{p}}\zeta_{k}^{\dagger}\zeta_{k}\punc.
\end{equation}
Note that it is possible to express \(\Phi_y\) in this way only because it can be written in terms of vertical dimers on a single row, as in \refeq{eq:yflux}. There is no corresponding operator for \(\Phi_x\), which necessarily involves dimers on different rows.

\subsection{Expectation values}

Expectation values in the classical ensemble can be expressed using appropriate operators acting in the Hilbert space. For the dimer occupation numbers, these are \cite{Wilkins2021}
\begin{equation}
\label{eq:dxy}
\begin{aligned}
d_{j,x} &= -\anis C_{j}C_{j+1} \\
d_{j,y} &= C_{j}^{\dagger}C_{j} \punc,
\end{aligned}
\end{equation}
for the horizontal bond between \(r_x = j\) and \(j+1\) and for the vertical bond at \(r_x = j\), respectively.

To calculate an expectation value involving observables at different rows \(l\), one writes out the operator exponential in \refeq{eq:partitionfunction} as a product of \(L_y\) copies of the single-row transfer matrix \(V\), and inserts operators for the observables at appropriate positions. For example, the correlation function for two observables, $O_1$ and $O_2$, in rows \(1 \le l_1 \le l_2 \le L_y\), is given by
\begin{equation}
\label{eq:correlatortrace}
\langle O_1(l_1)O_2(l_2) \rangle = \frac{1}{Z}\Tr\left[V^{L_{y}-l_2}\hat{O}_2 V^{l_2-l_1} \hat{O}_1 V^{l_1}\right] \punc ,
\end{equation}
where \(\hat{O}\) is the operator corresponding to the classical observable \(O\). When both \(l_1\) and \(l_2\) are even, this expression can easily be rewritten in terms of \(\ham\) using \refeq{eq:defineham}. To handle the case where either is odd, we define
\beq[eq:defineOl]
\hat{O}_{[l]} = \begin{cases}
\hat{O} & \text{for \(l\) even}\\
\ee^{-\ham}V^{-1}\hat{O}V\ee^{+\ham} & \text{for \(l\) odd,}
\end{cases}
\eeq
in terms of which
\begin{multline}
\label{eq:multirowev}
\langle O_1(l_1)O_2(l_2) \rangle\\{}= \frac{1}{Z}\Tr\left[\ee^{-\ham(L_{y}-l_2)}\hat{O}_{2[l_2]}\ee^{-\ham(l_2-l_1)}\hat{O}_{1[l_1]}\ee^{-\ham l_1}\right]\punc.
\end{multline}
Note that \(\hat{O}_{[l]}\) does not include the full \(l\) dependence (i.e., it is not an imaginary-time Heisenberg operator) and depends on \(l\) only through its parity. Clearly, \((O_1O_2)_{[l]}=O_{1[l]}O_{2[l]}\) and, because \(V\) is Hermitian,
\beq[eq:Oldagger]
\left(\hat{O}_{[l]}\right)^\dagger = \left(\hat{O}^\dagger\right)_{[l]}\punc.
\eeq

For the dimer operators in \refeq{eq:dxy}, we therefore have
\begin{align}
\label{eq:dxc}
d_{j[l],x} &= -\anis C_{j[l]}C_{j+1,[l]} \\
\label{eq:dyc}
d_{j[l],y} &= C_{j[l]}^{\dagger}C_{j[l]}\punc.
\end{align}
Using \(\ee^{-\ham}\zeta_k\ee^{+\ham} = \ee^{+\epsilon(k)}\zeta_k\) and
\beq[eq:onerowzeta]
\begin{aligned}
V^{-1}\zeta_{k}^*V &= - \ee^{-\epsilon(k)}\zeta_{k-\pi}^{\dagger}\\
V^{-1}{\big(\zeta_{k}^\dagger\big)}^*V &= - \ee^{\epsilon(k)}\zeta_{k-\pi}\punc,
\end{aligned}
\eeq
together with the complex conjugate of \refeq{eq:czeta}, we find
\begin{multline}
\label{eq:Cjl}
C_{j[l]} = \sqrt{\frac{2}{L_x}}\ee^{-\ii(-1)^l \pi/4}\sum_{k\in\mathbb{K}_p} \ee^{\ii k j}\cos\theta_k \\
{} \times \begin{cases}
 \zeta_{k} & \text{for \(j+l\) even} \\
(-1)^l\zeta_{-k}^\dagger & \text{\phantom{for} \(j+l\) odd.}
\end{cases}
\end{multline}

\subsection{Monomer distribution function}
\label{monomerdistributionfunction}

The monomer distribution function \(G\sub{m}\), related to the entropic interaction between a pair of monomers (empty sites) \cite{Powell2013}, is defined by
\beq
G\sub{m}(\rv_1,\rv_2) = \frac{1}{Z}\sum_{c\in\goC(\rv_1,\rv_2)} \anis^{N_{x}}\punc ,
\eeq
where \(\goC(\rv_1,\rv_2)\) is the set of all configurations with monomers at sites \(\rv_1\) and \(\rv_2\).

In the transfer-matrix formalism, an ensemble containing monomers can be produced by inserting spin-lowering operators \(\sigma^-\) into the original partition function. To see this, suppose the vector \(\lvert l_0\rangle\) gives all possible configurations on row \(l_0\) of vertical bonds, with appropriate weights. Then \(\sigma_j^- V^{l-l_0}\lvert l_0\rangle\) gives all configurations of row \(l\) where bond \(j\) is empty but where the intervening rows are consistent with it being occupied. The resulting configurations and their weights are exactly those with a monomer in column \(j\) of the row of lattice sites between rows \(l-1\) and \(l\).

The monomer distribution function is therefore given by \cite{Wilkins2021}
\beq
G\sub{m}(\rv_1,\rv_2) = \left\langle \sigma_{j_1}^{-}(l_1)\sigma_{j_2}^{-}(l_2) \right\rangle\punc,
\eeq
where the right-hand side is defined as in \refeq{eq:correlatortrace}. In the case where the monomers are in the same row and at columns \(j\) and \(j+X\) respectively, this can be expressed as
\begin{equation}
G\sub{m}(\rv_1,\rv_2) = -\left\langle C_{j} \ee^{\ii\pi \sum_{i=j+1}^{j+X-1}C_i^\dagger C_i} C_{j+X} \right\rangle\punc,
\label{eq:monomerfermions}
\end{equation}
using the inverse of the Jordan--Wigner mapping, \refeq{eq:JW1}. This expectation value can be calculated in the microscopic theory using Wick's theorem \cite{Wilkins2021}.

\section{Derivation of field theory}
\label{derivationoffieldtheory}

In this section, we derive an effective field theory description of the dimer model using bosonization, starting from the transfer-matrix solution.

\subsection{Bosonized Hamiltonian and action}

\subsubsection{Linearized theory}
\label{linearizedtheory}

\begin{figure}
\begin{center}
\includegraphics[width=\columnwidth]{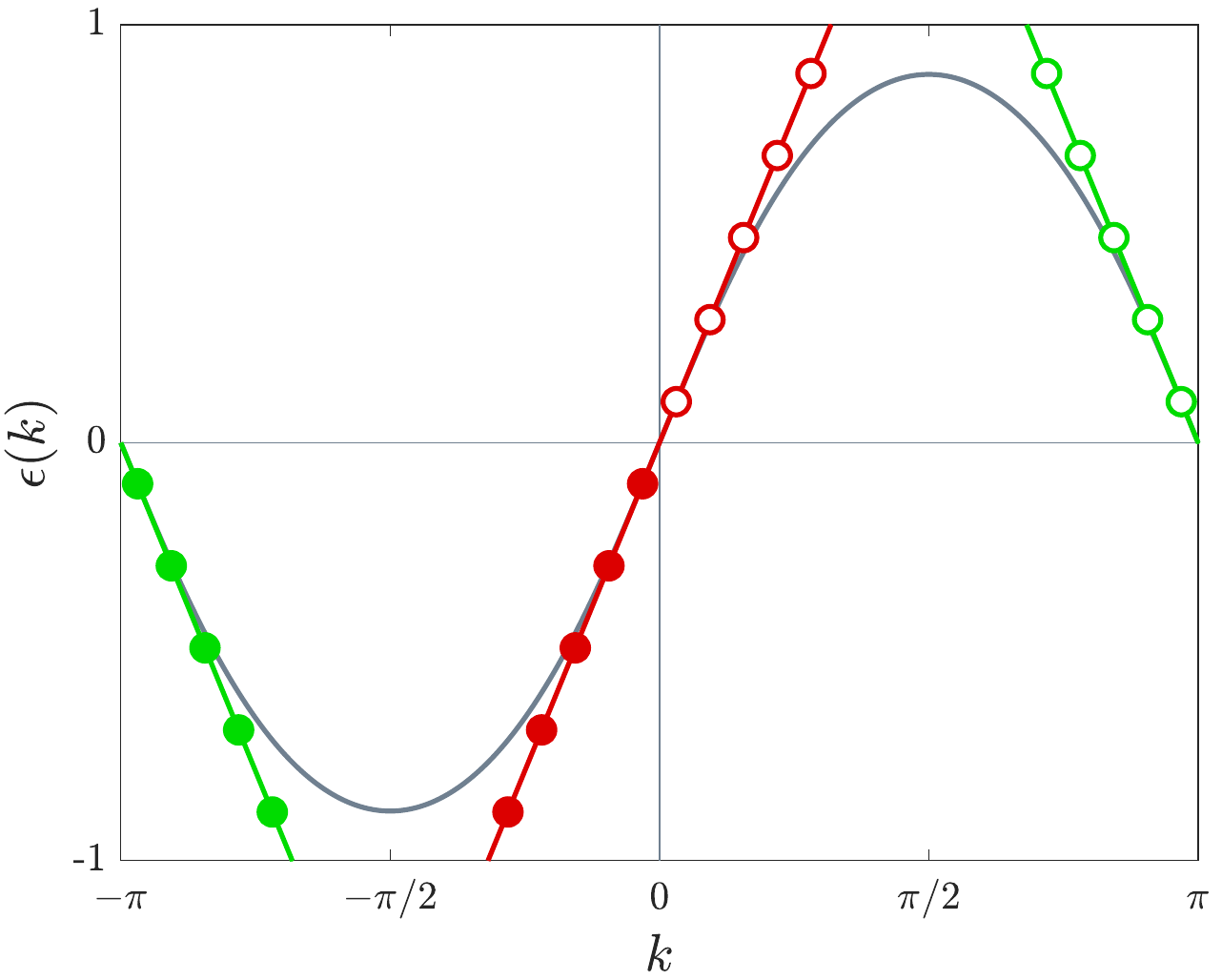}
\caption{The dispersion $\epsilon(k)$ (gray) is linearized around the two Fermi points at $k_+ = 0$ and $k_- = \pi$; the right- (red) and left- (green) moving branches are then  extended to infinity. Filled and empty circles denote single-fermion states that are occupied and empty, respectively, in the ground state. (The dispersion is plotted for the isotropic case \(\anis=1\) and we have set \(p=0\), so the \(k\) values are half-integer multiples of \(2\pi/L_x\).)}
\label{fig:linearization}
\end{center}
\end{figure}

The free-fermion Hamiltonian, \refeq{eq:ham2}, has dispersion $\epsilon(k)$, as shown in \reffig{fig:linearization}. There are two Fermi points \(k_\pm\), i.e., points where $\epsilon(k_\pm) = 0$, at
\begin{equation}
\label{eq:FP} \qquad k_+ = 0 \qquad \text{and} \qquad k_- = \pi \punc ,
\end{equation}
which we refer to as the right and left Fermi points respectively. The Fermi velocities at these points are \(\epsilon'(k_\pm) = \pm \anis\).

Our goal is to construct a linearized theory that exactly reproduces long-wavelength properties, such as long-distance correlation functions, of the microscopic model. To do so, we linearize the dispersion around the two Fermi points and extend each branch to infinity (see \reffig{fig:linearization}). This modifies the dispersion away from the Fermi points and also adds an infinite number of new single-particle modes at high energy. On physical grounds, we expect the long-wavelength properties to be unaffected, provided we consider only quantities to which the extra modes do not make spurious contributions.

To describe the modes of the linearized theory, we define right- and left-moving fermion annihilation operators, \(R_k\) and \(L_k\), and corresponding real-space operators
\beq[eq:fs]
\begin{aligned}
R(x) &= \frac{1}{\sqrt{L_{x}}}\sum_{k=-\infty}^{\infty}\ee^{+\ii kx}R_{k}\\
L(x) &= \frac{1}{\sqrt{L_{x}}}\sum_{k=-\infty}^{\infty}\ee^{-\ii kx}L_{k}\punc.
\end{aligned}
\eeq
The set of \(k\) values included in each sum is the infinite extension of \(\mathbb{K}_p\): all half-integer (integer) multiples of \(2\pi/L_x\) for \(p=0\) (resp., \(p=1\)). This implies the boundary condition
\begin{equation}
R(x+L_x) = -(-1)^p R(x)
\end{equation}
and the same for \(L(x)\). Note that the real-space operators are defined in the continuum rather than on the lattice, since there is no Brillouin zone in the linearized theory.

These operators are related to the microscopic fermions \(\zeta_k\) by
\begin{equation}
\label{eq:zetaLR}
\zeta_{k} \approx \begin{cases}
R_{k - k_+} & \text{for \(k\) near \(k_+\)}\\
L_{k_- - k} & \text{\phantom{for} \(k\) near \(k_-\),}
\end{cases}
\end{equation}
or equivalently to the real-space fermions
\beq[eq:psix]
\psi_{j} = \frac{1}{\sqrt{L_{x}}}\sum_{k}\ee^{\ii kj}\zeta_{k}
\eeq
by
\beq[eq:psiLR]
\psi_{j} \approx \ee^{\ii k_+j}R(x_j) + \ee^{\ii k_-j}L(x_j)\punc,
\eeq
where \(x_j\) is the position of site \(j\). Expressions such as \refeqand{eq:zetaLR}{eq:psiLR} relate operators in the microscopic and linearized theories, which have different Hilbert spaces. They should be understood as equivalences between the two, rather than (approximate) operator identities.

When constructing corresponding relations for other operators such as the Hamiltonian, it is necessary to avoid spurious contributions from the additional modes in the linearized theory. This can be achieved by putting all fermion operators into normal order (see \refapp{normalordering}) before applying \refeq{eq:zetaLR} or \refeq{eq:psiLR}. For bilinears, normal ordering simply amounts to subtracting the ground-state expectation value, and so the Hamiltonian, \refeq{eq:ham2}, can be written as
\begin{equation}
\label{eq:hamno}
\ham_p = E_{0}(p) + \sum_{k}\epsilon(k)\no{\zeta_{k}^{\dagger}\zeta_{k}}\punc,
\end{equation}
where \(\no{\quad}\) denotes normal ordering.
Here, \(E_0(p)\) is the ground-state energy in the sector with parity \(p\), which obeys \cite{Wilkins2021}
\begin{equation}
\label{eq:E01}
E_0(1) = E_0(0) + \frac{\pi\anis}{2L_x} + O\left(L_x^{-3}\right) \punc.
\end{equation}
One similarly finds, using \refeq{eq:fluxzeta},
\beq[eq:fluxzeta2]
\Phi_{y} = p + \sum_{k}\no{\zeta_{k}^{\dagger}\zeta_{k}}\punc.
\eeq
(Our convention for fermion normal-ordering, defined in \refapp{fermions}, uses as reference state the ground state with \(\Phi_y = +1\) when \(p=1\).)

Using \refeq{eq:zetaLR} and expanding $\epsilon(k)$ to leading order around the Fermi points then gives
\begin{equation}
\label{eq:firstorder}
\ham_p = E_{0}(p) +\anis\sum_{k}k\left(\no{R_{k}^{\dagger}R_{k} + L_{k}^{\dagger}L_{k}}\right)  \punc ,
\end{equation}
or, transforming to real space by inverting \refeq{eq:fs},
\begin{equation}
\ham_p = E_{0}(p) + \frac{\anis}{\ii}\int_{0}^{L_{x}} \!\dd x \, \left[\no{R^{\dagger}(x)\partial_{x}R(x) - L^{\dagger}(x)\partial_{x}L(x)}\right]  \punc .
\end{equation}
Here and in the following, we use the same symbols for operators in the microscopic and linearized theories (or, in other words, write these relationships using ``\(=\)'' rather than ``\(\approx\)''), since there is no risk of ambiguity. The corresponding expression for the flux is
\beq[eq:PhiyRL]
\Phi_y = p + \int_0^{L_x}\!\dd x\,\left[ \no{R^{\dagger}(x)R(x) + L^{\dagger}(x)L(x)}\right]\punc.
\eeq

\subsubsection{Bosonization and path integral}
\label{action}

\renewcommand{\arraystretch}{2.5}
\newcommand{\eqn}[1]{\;\(\displaystyle #1\)\;}
\begin{table*}
\begin{center}
\begin{tabular}{r@{\({}={}\)}l@{\extracolsep{1in}}c}
\hline\hline
\multicolumn{1}{r@{\({}\leftrightarrow{}\)}}{{Fermions\,}}& \multicolumn{1}{l}{Bosons} & {Equation} \\ \hline
\eqn{\no{R^{\dagger}R + L^{\dagger}L}} & \eqn{\partial_{x}\phi - \frac{1}{L_x}p} & \refeq{eq:dxphi} \\
\eqn{\no{R^{\dagger}R - L^{\dagger}L}} & \eqn{-\frac{1}{\pi}\Pi} & \refeq{eq:Pi} \\
\eqn{\frac{1}{2\ii}\left[\no{R^\dagger\partial_{x}R - (\partial_{x}R^{\dagger})R - L^\dagger\partial_{x}L + (\partial_{x}L^{\dagger})L}\right]} & \eqn{\frac{\pi}{2}\left[\bno{(\partial_{x}\phi)^{2}+\frac{1}{\pi^2}\Pi^{2}}\right] - \frac{\pi}{2L_x^2}p
} & \refeq{eq:hamnoint} \\
\eqn{R^{\dagger}L + L^{\dagger}R} & \eqn{\frac{1}{\pi a}\sin (2\pi \phi)} & \refeq{eq:RLsin} \\
\eqn{R^{\dagger}L - L^{\dagger}R} & \eqn{\frac{\ii}{\pi a}\cos (2\pi \phi)} & \refeq{eq:RLcos} \\
\hline\hline
\end{tabular}
\caption{Bosonization dictionary: identities, derived in \refapp{bosonization}, that relate bilinears in right- and left-moving fermion fields, \(R = \Psi_+(x)\) and \(L = \Psi_-(x)\), to expressions containing the boson field $\phi$ and its canonically conjugate momentum $\Pi$. The fermion fields obey boundary conditions \(\Psi_\pm(x+L_x) = -(-1)^p\Psi_\pm(x)\) with \(p=0\) or \(1\), and fermion normal ordering, denoted \(\no{\quad}\), is defined with respect to the ground state subject to this boundary condition. Boson normal ordering is denoted \(\bno{\quad}\). The constant \(a\) is a short-distance cutoff with dimensions of length that arises in the bosonization formalism and which we use to regularize the bosonic field theory (see \refapp{regularization}).}
\label{tab:bosonizationdictionary}
\end{center}
\end{table*}
\renewcommand{\arraystretch}{1}

The right- and left-moving fermions can be expressed, according to the bosonization identity \cite{vonDelft1998}, in terms of a pair of bosonic operators \(\phi(x)\) and \(\Pi(x)\). We provide full details of this mapping in \refapp{bosonization}, but \reftab{tab:bosonizationdictionary} summarizes the most important results. [These are operator identities, in contrast to relations such as \refeq{eq:zetaLR}.]

The operators \(\phi(x)\) and \(\Pi(x)\) are Hermitian and obey \([\phi(x),\phi(x')] = [\Pi(x),\Pi(x')] = 0\) and
\beq{}
[\phi(x),\Pi(x')] = \ii \delta_a(x-x')
\eeq
for \(0 \le x,x' < L_x\). Here, \(\delta_a\) is a nascent delta function [specifically, a Lorentzian; see \refeq{eq:Lorentzian}] whose width is \(a\), a short-distance cutoff that arises in the bosonization formalism \cite{vonDelft1998} and which we use to regularize the field theory (see \refapp{regularization}). In the limit \(a \rightarrow 0\), \(\delta_a\) becomes a Dirac delta function, and so we refer to \(\Pi(x)\) as the canonically conjugate momentum for the field \(\phi(x)\). (Note that \(a\) is \emph{not} the lattice constant, which we have set to \(1\).)

As we show in \refapp{bosonizationidentity}, \(\phi(x)\) is defined only up to a global integer shift and has a jump across the periodic boundaries. To be precise, \refeqand{eq:PhiyRL}{eq:dxphi} give
\beq[eq:phiydphi]
\Phi_y = \int_0^{L_x}\dd x\,\partial_x \phi\punc,
\eeq
and so \(\phi(x)\) obeys the boundary condition [equivalent to \refeq{eq:phiaperiodic2}]
\beq[eq:phibc]
\phi(L_{x}) = \phi(0) + \Phi_y\punc.
\eeq
For the Hamiltonian, using \refeq{eq:hamnoint} and integration by parts gives
\begin{equation}
\label{eq:hamthetaphi}
\ham_p = E_0(0) + \frac{\anis \pi}{2}\int_{0}^{L_{x}}\!\dd x \left[\bno{(\partial_{x}\phi)^{2} + \frac{1}{\pi^2}\Pi^{2}}\right] \punc ,
\end{equation}
where \(\bno{\quad}\) denotes boson normal ordering (see \refapp{bosons}). Note that, using \refeq{eq:E01}, the dependence on parity sector \(p\) cancels at least to order \(L_x^{-2}\), and we therefore omit \(p\) in the following.

Using the standard mapping to a path integral for bosonic fields  \cite{Sachdev2011}, the partition function, \refeq{eq:partitionfunction2}, may be rewritten as
\beq[eq:partitionfunction3]
Z = \int \mathcal{D}\phi \, \ee^{-S[\phi]} \punc ,
\eeq
where the action functional is\footnote{The boson normal ordering must be removed before deriving the path integral, but this merely adds an unimportant constant \(\propto a^{-2}\).}
\begin{equation}
\label{eq:sphi}
S[\phi] = \frac{\anis\pi}{2}\int_0^{L_x} \!\dd x \int_0^{L_y} \!\dd y \, \left[(\partial_{x}\phi)^{2} + \frac{1}{\anis^2}(\partial_{y}\phi)^{2}\right] \punc.
\end{equation}
We label the imaginary-time dimension in the path integral \(y\), because the transfer matrix effects evolution in the \(y\) direction of the original dimer model; the effective inverse temperature is \(L_y\) [see \refeq{eq:partitionfunction2}].

The integral in \refeq{eq:partitionfunction3} is over real field configurations \(\phi(x,y)\), with \(0 \le x \le L_x\) and \(0 \le y \le L_y\). Due to the boundary condition \refeq{eq:phibc}, it includes paths where \(\phi(L_x,y)-\phi(0,y)\) is given by an integer, identified as the \(y\) component of the flux, \(\Phi_y\).

Since the operator \(\phi(x)\) is defined only up to global integer shifts, so are the fields appearing in the path integral. Fixing the value of \(\phi\) at any one point \((x,y)\) allows it to be fixed everywhere throughout a simply connected region, but not around the periodic boundaries in the \(y\) direction. The path integral should therefore also include configurations where \(\phi(x,L_y)-\phi(x,0)\) is an integer.\footnote{An identical argument applies for the path integral of a single \(\mathrm{O}(2)\) quantum rotor \cite{Sachdev2011}.} As expected by symmetry, and as we show explicitly in \refsec{dimeroperatorsinpathintegral} [see \refeq{eq:Phixphi}], this discontinuity determines the \(x\) component of the flux, \(\Phi_x\).

The field theory defined by \refeqand{eq:partitionfunction3}{eq:sphi} -- which is simply a free, massless real field theory in 2D -- is identical to the ``height theory'' proposed to describe the dimer model \cite{Fradkin2004,Fradkin2013,Papanikolaou2007,Tang2011}, including the invariance under global shifts and the relationship between the flux and the boundary conditions on the field. We return to this point in \refsec{dimeroperatorsinpathintegral} where we relate the field \(\phi\) to the dimer observables.

\subsection{Dimer occupation numbers}
\label{dimeroccupationnumbers}

The field theory for the dimer model, \refeqand{eq:partitionfunction3}{eq:sphi}, is only useful when paired with expressions that relate the microscopic degrees of freedom, i.e., dimers, to the field $\phi$, so that one can calculate observables. In this section we derive expressions for the dimer occupation numbers on vertical and horizontal bonds in terms of $\phi$.

\subsubsection{Bosonized operators}

The first step is to write the expressions for \(d_{j[l],x}\) and \(d_{j[l],y}\), \refeqand{eq:dxc}{eq:dyc}, in normal order. Since these operators are bilinears in fermions, this simply involves subtracting their expectation values in the reference state \(\lvert 0 \rangle\), the ground state of the free-fermion Hamiltonian \(\ham_p\) in parity sector \(p\) (see \refapp{fermions}). It is in fact more convenient to express them in terms of expectation values in the \emph{global} ground state of \(\ham\), which is equal to the ground state in the sector with \(p=0\). Denoting such expectation values by \(\langle \quad \rangle\), we find, by a similar calculation to that for \refeq{eq:E01},
\begin{align}
\langle 0 \rvert d_{j[l],x} \lvert 0 \rangle &= \langle d_x \rangle\\
\langle 0 \rvert d_{j[l],y} \lvert 0 \rangle &= \langle d_y \rangle + p\frac{(-1)^{j+l}}{L_x}\punc,
\end{align}
up to corrections of higher order in \(L_x^{-1}\). By symmetry, \(\langle d_x \rangle\) and \(\langle d_y \rangle\) are exactly independent of position, which is not true of \(\langle 0 \rvert d_{j[l],y} \lvert 0 \rangle\) for \(p=1\), since \(\Phi_y \neq 0\) reduces the symmetry.\footnote{For completeness, their values are
\[
\langle d_x \rangle = \frac{\arctan{\anis}}{\pi} \qquad
\langle d_y \rangle = \frac{\arctan{(1/\anis)}}{\pi}\punc,
\]
in the thermodynamic limit \cite{Fisher1963,Wilkins2021}.} We therefore have
\begin{align}
d_{j[l],x} - \langle d_{x} \rangle &= -\anis\no{C_{j[l]}C_{j+1,[l]}} \\
d_{j[l],y} - \langle d_y \rangle &= \no{C_{j[l]}^{\dagger}C_{j[l]}} + p\frac{(-1)^{j+l}}{L_x} \punc .
\label{eq:dylj}
\end{align}
While the term of order \(L_x^{-1}\) in \refeq{eq:dylj} has no effect on dimer correlations in the thermodynamic limit, it is required to give the correct expression for the flux in \refsec{dimeroperatorsinpathintegral}.

Next, we express these in terms of right- and left-moving fermions \(R(x)\) and \(L(x)\), by linearizing the transformation between $C_{j[l]}$ and $\zeta_{k}$, \refeq{eq:Cjl}. After inserting \refeq{eq:zetaLR}, shifting the summation indices, and extending both sums over $k$ to infinity, we obtain
\begin{multline}
\label{eq:CjlRL}
C_{j[l]} = \sqrt{\frac{2}{L_{x}}}\ee^{-\ii(-1)^l \pi/4}\sum_{k}\ee^{\ii k j} \\
{}\times \begin{cases}
\cos\theta_{k}R_{k} + (-1)^l\sin\theta_{k}L_{-k} & \text{for \(j+l\) even}\\
(-1)^l\cos\theta_{k}R_{-k}^{\dagger} - \sin\theta_{k}L_{k}^{\dagger} & \text{\phantom{for} \(j+l\) odd.}
\end{cases}   
\end{multline}
By assumption, the only important contributions to the sum have small \(k\), and so we replace \(\theta_k\) by \(\theta_0 = \frac{\pi}{4}\). The sums can then be identified with real-space fermions using \refeq{eq:fs}, which gives
\beq[eq:clr]
C_{j[l]} = \ee^{-\ii (-1)^l \pi/4}\times\begin{cases}
R + (-1)^lL & \text{for \(j+l\) even}\\
(-1)^lR^\dagger - L^\dagger & \text{\phantom{for} \(j+l\) odd.}
\end{cases} 
\eeq
Like those in \refsec{linearizedtheory}, this equation relates an operator in the microscopic theory to the equivalent operator in the linearized theory. Here and in the expressions that follow, the continuum fields on the right-hand side are understood to be evaluated at the position \(x_j\) of site \(j\).

Inserting this result into \refeq{eq:dylj} gives, for the dimer occupation number on vertical bonds,
\begin{multline}
\label{eq:lrres}
d_{j[l],y} - \langle d_y\rangle
= (-1)^{j+l}\no{\hc{R}R + \hc{L}L}
\\+ (-1)^l\left(\hc{L}R + \hc{R}L\right)
\punc .
\end{multline}
This can be bosonized using the results of \reftab{tab:bosonizationdictionary}, giving
\begin{equation}
\label{eq:dyfinaloperator}
d_{j[l],y} - \langle d_y\rangle = (-1)^{j+l}\partial_x\phi + \frac{(-1)^l}{\pi a}\sin(2\pi\phi) \punc .
\end{equation}
Similarly, for the dimer occupation number on horizontal bonds, we find
\begin{multline}
d_{j[l],x} - \langle d_x\rangle = -\ii\anis\bigg[(-1)^{j+l}\no{R^{\dagger}R - L^{\dagger}L} \\{}+ (-1)^j (R^{\dagger}L - L^{\dagger}R) \bigg] \punc ,
\end{multline}
and bosonization gives
\begin{equation}
\label{eq:dx}
d_{j[l],x} - \langle d_{x}\rangle = \frac{\ii\anis}{\pi}(-1)^{j+l}\Pi + \frac{\anis}{\pi a}(-1)^{j}\cos(2\pi\phi) \punc .
\end{equation}

These leading-order expressions for \(d_{j[l],x}\) and \(d_{j[l],y}\) result from replacing \(\theta_k\) in \refeq{eq:CjlRL} by its value at \(k=0\). One can instead expand in a Taylor series around this point, which gives terms in \refeq{eq:clr} involving derivatives of the fermion fields \(R\) and \(L\). These terms can also be bosonized  \cite{WilkinsThesis}, giving higher-order terms in \refeqand{eq:dyfinaloperator}{eq:dx} and hence corrections to the asymptotic dimer--dimer correlations.

\subsubsection{Dimer operators in path-integral formalism}
\label{dimeroperatorsinpathintegral}

Applying the same mapping that led from the partition function in the form \refeq{eq:partitionfunction2} to the path integral \refeq{eq:partitionfunction3}, we can write any expectation value as an average over the ensemble of field configurations. For example, the correlation function in \refeq{eq:multirowev} becomes
\begin{equation}
\label{eq:pathaverage}
\langle O_1(l_1)O_2(l_2) \rangle = \frac{1}{Z}\int \mathcal{D}\phi \, \ee^{-S[\phi]}\widetilde{O_{1,[l_1]}}(l_1)\widetilde{O_{2,[l_2]}}(l_2)\punc,
\end{equation}
where \(\widetilde{O}(l)\) is a function of \(\phi\) and its derivatives evaluated at \(y=l\). In cases where the operator \(\hat{O}\) can be expressed in terms of only the field operator \(\phi\), \(\widetilde{O}(l)\) is simply given by the corresponding function of the field configuration \(\phi\). For the operator \(\Pi\), which is not diagonal in the basis states used to construct the path integral, the appropriate function is given by the ``equation of motion'' $\widetilde{\Pi} = \frac{\ii\pi}{\anis}\partial_{y}\phi$.

Mapping \refeqand{eq:dyfinaloperator}{eq:dx} to the path-integral formulation therefore gives (omitting \(\widetilde{\;\;}\) to simplify the notation)
\begin{equation}
\label{eq:dyfinalfield}
d_{y}(\rv) - \langle d_y\rangle = (-1)^{r_x+r_y}\dx\phi + \frac{1}{\pi a}(-1)^{r_y}\sin(2\pi\phi)
\end{equation}
and\footnote{As usual with off-diagonal operators in the path integral, this expression does not give the correct expectation value for an operator such as \([d_x(\rv)]^2\). Here, this operator does not give the correct expectation value for the dimers anyway \cite{Wilkins2021}.}
\begin{equation}
\label{eq:dxfinal}
d_{x}(\rv) - \langle d_x\rangle = -(-1)^{r_x+r_y}\partial_{y}\phi + \frac{\anis}{\pi a}(-1)^{r_x}\cos(2\pi\phi)\punc.
\end{equation}
These expressions, along with the action in \refeq{eq:sphi}, are the main results of the bosonization calculation.

The relations in \refeqand{eq:dyfinalfield}{eq:dxfinal} are exactly equivalent to those proposed to relate dimer observables to a height field describing long-wavelength properties of the dimer model \cite{Fradkin2004,Fradkin2013,Papanikolaou2007,Tang2011,Alet2016}. In these previous works, their form was inferred based on the microscopic definition of the height and symmetry considerations (and by analogy with the bosonized form of the spin and charge density in a Luttinger liquid \cite{Fradkin2013}), while the coefficients, along with those in the action, were fixed using exact asymptotic results for the dimer correlations. Here, we have explicitly constructed the long-wavelength theory describing the model and shown that it contains a single real field \(\phi\), related to the dimer observables according to \refeqand{eq:dyfinalfield}{eq:dxfinal} and governed by the action \refeq{eq:sphi}, with no undetermined parameters. By comparing \refeqand{eq:dyfinalfield}{eq:dxfinal} for \(\anis=1\) with the corresponding expressions in \refcite{Fradkin2004,Papanikolaou2007}, we conclude that our field \(\phi\) is related to the continuum height fields of the latter by \(\phi = -\frac{1}{2\pi}\phi\sub{\cite{Papanikolaou2007}} = -\frac{1}{4}h\sub{\cite{Fradkin2004}}\) (where the subscripts are reference numbers).\footnote{\label{footnote:cutoff}Note that \refeqand{eq:dyfinalfield}{eq:dxfinal} include explicit factors of the short-distance regularization parameter \(a\). As we show in \refsec{dimerdimercorrelations}, this cancels in observables such as dimer--dimer correlation functions. Expressions for the dimers in terms of the height that do not explicitly include a cutoff, such as in \refcites{Fradkin2004,Papanikolaou2007,Fradkin2013,Tang2011}, assume an appropriate choice of cutoff in terms of the lattice spacing (which is set to \(1\) here and in the cited references).}

From \refeqand{eq:dyfinalfield}{eq:dxfinal}, we can determine the transformation properties of \(\phi\) under the symmetries of the microscopic model. Under \(\mathbb{T}_x\) and \(\mathbb{T}_y\), translation by one lattice spacing in the \(x\) and \(y\) directions respectively, the right-hand sides of \refeqand{eq:dyfinalfield}{eq:dxfinal} should transform trivially, which requires
\begin{align}
\label{eq:Tphi}
\phi &\xrightarrow{\mathbb{T}_x} \frac{1}{2} - \phi & \phi &\xrightarrow{\mathbb{T}_y} - \phi\punc.
\end{align}

When \(\anis=1\), there is also symmetry under a fourfold rotation \(\mathbb{R}\), which we define as rotation by \(\frac{\pi}{2}\) counterclockwise around \(\rv = \zerov\). Under this transformation,\footnote{The bond joining sites \(\mathbb{R}^{-1}\rv - \deltav_y\) and \(\mathbb{R}^{-1}\rv\), whose dimer occupation number is \(d_y(\mathbb{R}^{-1}\rv - \deltav_y)\), is mapped by \(\mathbb{R}\) to the bond joining \(\rv\) and \(\rv + \deltav_x\).}
\begin{align}
d_x(\rv) &\rightarrow d_y(\mathbb{R}^{-1}\rv - \deltav_y)\\
d_y(\rv) & \rightarrow d_x(\mathbb{R}^{-1}\rv)\punc,
\end{align}
and so
\beq[eq:Rphi]
\phi \xrightarrow{\mathbb{R}} \phi + \frac{1}{4}\punc.
\eeq
Note that the similarity between the expressions for the vertical and horizontal dimers, \refeqand{eq:dyfinalfield}{eq:dxfinal}  respectively, reflects the fact that the symmetry (for \(\anis=1\)) between \(x\) and \(y\), hidden in the transfer-matrix formalism, is manifest in the bosonized path integral.

Transformation properties equivalent to \refeqand{eq:Tphi}{eq:Rphi} were determined in \refcite{Alet2006b} based on the microscopic definition of the height; comparison with these results implies \(\phi = h\sub{\cite{Alet2006b}} - \frac{1}{8}\). (The rotation considered in \refcite{Alet2006b} is about a plaquette center and can be expressed as \(\mathbb{R}' = \mathbb{T}_y\mathbb{R}\). Reference~\cite{Wilkins2020} uses the definitions of \refcite{Alet2006b}.)

Following \refcite{Alet2006b}, one can also consider the consequences of locality. The dimer occupation numbers \(d_\mu(\rv)\) are given by local functions of the field \(\phi(\rv)\) and its derivatives, but the derivative terms in \refeqand{eq:dyfinalfield}{eq:dxfinal} imply that \(\phi(\rv)\) depends nonlocally on the dimer configuration. Nonetheless, any rearrangement of the dimers can only change \(d_x\) and \(d_y\) by integers, and so moving dimers within a region far from \(\rv\) can only shift \(\phi(\rv)\) by an integer. This implies that the action must be invariant under integer shifts of \(\phi(\rv)\), for any microscopic model that is local in the dimer degrees of freedom, as previously determined \cite{Alet2006b} by considering the microscopic heights. Note that the redundancy under integer shifts arises naturally from the bosonization procedure, as described in \refsec{action}.

Finally, using \refeqand{eq:dyfinalfield}{eq:dxfinal} we can confirm the relationship between the flux \(\Phiv\) and the boundary conditions on \(\phi\) stated at the end of \refsec{action}. We start from \refeq{eq:yflux}, which gives \(\Phi_y\) as the sum of \((-1)^{r_x+r_y}d_{y}(\rv)\) over any row \(r_y\) of the lattice, and use \refeq{eq:dyfinalfield}. Since \(\phi\) is a slowly varying function of \(\rv\) and \(\langle d_y \rangle\) is independent of \(\rv\), we can drop rapidly oscillating terms and replace the sum by an integral, leaving
\beq
\Phi_y = \int_0^{L_x} \dd x\, \partial_x \phi = \phi(L_x, r_y) - \phi(0, r_y)\punc,
\eeq
as claimed above. Similarly, \refeqand{eq:xflux}{eq:dxfinal} give
\beq[eq:Phixphi]
\Phi_x = -\int_0^{L_y} \dd y\, \partial_y \phi = -\phi(r_x, L_y) + \phi(r_x, 0)\punc,
\eeq
for any \(r_x\).

\subsubsection{Dimer--dimer correlations}
\label{dimerdimercorrelations}

To check our results for the action, \refeq{eq:sphi}, and the dimer occupation numbers, \refeqand{eq:dyfinalfield}{eq:dxfinal}, we use them to calculate the asymptotic behavior of the dimer--dimer correlation function,
\begin{equation}
\label{eq:gxx2}
G^{\mu\nu}(\Rv) = \langle d_{\mu}(\rv+\Rv)d_{\nu}(\rv)\rangle - {} \\ \langle d_{\mu}(\rv+\Rv)\rangle\langle d_{\nu}(\rv)\rangle \punc,
\end{equation}
for large separation $\Rv = (X,Y)$, and compare with results calculated for the microscopic lattice model \cite{Fisher1963,Wilkins2021}.

As we show in \refapp{regularization}, the field theory has correlation function
\begin{equation}
\label{eq:asymptoticcorrelations}
\left\langle \left[\phi(\rv+\Rv) - \phi(\rv) \right]^2\right\rangle =\frac{1}{\pi^2} \log \frac{\lvert \tilde{\Rv} \rvert}{a} + O\left(\frac{\lvert\tilde{\Rv}\rvert}{a}\right)^{-1}
\end{equation}
for \(a \ll \lvert \Rv \rvert \ll L_x,L_y\), where \(\tilde{\Rv}=(X,\anis Y)\). [The average \(\langle\quad\rangle\) is taken in the ensemble of field configurations, defined as on the right-hand side of \refeq{eq:pathaverage}.]

Differentiating this, we find
\begin{equation}
\left\langle\partial_{\mu}\phi(\rv+\Rv)\,\partial_{\nu}\phi(\rv)\right\rangle = \frac{\anis_\mu\anis_\nu}{2\pi^2}\frac{\lvert\tilde{\Rv}\rvert^2\delta_{\mu\nu} - 2\tilde{R}_\mu\tilde{R}_\nu}{\lvert\tilde{\Rv}\rvert^4} \punc ,
\end{equation}
where \(\anis_x=1\) and \(\anis_y=\anis\), while the result $\langle \ee^{\ii A}\rangle = \ee^{-\frac{1}{2}\langle A^{2}\rangle}$, for a Gaussian random variable $A$ with zero mean, gives
\beq
\left\langle\ee^{2\pi\ii\big[\phi(\rv+\Rv)-\phi(\rv)\big]}\right\rangle =\frac{a^{2}}{\lvert\tilde{\Rv}\rvert^2}
\eeq
and \(\langle\ee^{2\pi\ii[\phi(\rv+\Rv)+\phi(\rv)]}\rangle =0\). We therefore have
\begin{multline}
\label{eq:sinecorrelator}
\Big\langle \cos\big(2\pi\phi(\rv+\Rv)\big)\cos\big(2\pi\phi(\bm{\rv})\big)\Big\rangle \\= \Big\langle \sin\big(2\pi\phi(\rv+\Rv)\big)\sin\big(2\pi\phi(\bm{\rv})\big)\Big\rangle = \frac{a^{2}}{2\lvert\tilde{\Rv}\rvert^2}
\end{multline}
and
\begin{equation}
\Big\langle \cos\big(2\pi\phi(\rv+\Rv)\big)\sin\big(2\pi\phi(\bm{\rv})\big)\Big\rangle = 0\punc.
\end{equation}

Substituting \refeqand{eq:dyfinalfield}{eq:dxfinal} into \refeq{eq:gxx2} and using these results, we find
\begin{align}
G^{xx}(\Rv) &\approx (-1)^X\frac{\anis^2}{\pi^2}\frac{1}{\lvert\tilde{\Rv}\rvert^4}\times\begin{cases}
X^2 & \text{\(Y\) even}\\
(\anis Y)^2 & \text{\(Y\) odd}
\end{cases}\\
G^{yy}(\Rv) &\approx (-1)^Y\frac{1}{\pi^2}\frac{1}{\lvert\tilde{\Rv}\rvert^4}\times\begin{cases}
(\anis Y)^2 & \text{\(X\) even}\\
X^2 & \text{\(X\) odd}
\end{cases}\\
G^{xy}(\Rv) &\approx (-1)^{X+Y}\frac{\anis^2}{\pi^2}\frac{ XY}{\lvert\tilde{\Rv}\rvert^4}\punc.
\end{align}
These are consistent with \refcite{Wilkins2021}, where the same asymptotic results were found by calculating correlation functions in the microscopic lattice model using the transfer matrix and then taking the large-separation limit.

As required for any observable, these correlation functions are independent of the regularization parameter \(a\). While the expressions for the dimer observables, \refeqand{eq:dyfinalfield}{eq:dxfinal}, contain \(a\) explicitly, this is compensated by the implicit \(a\) dependence of the distribution of fields \(\phi(\rv)\) implied by \refeq{eq:asymptoticcorrelations}.

\subsection{Monomers}
\label{monomers}

The monomer distribution function \(G\sub{m}\), which describes the dimer ensemble in the presence of a pair of monomers, is defined in \refsec{monomerdistributionfunction}. Because of the chain of operators in \refeq{eq:monomerfermions}, we are not able to bosonize it using the same approach as in \refsec{dimeroccupationnumbers}. Instead, we construct an operator in the continuum theory that corresponds to the spin-lowering operator \(\sigma_j^-\) and confirm that it reproduces the correct leading-order behavior for the monomer distribution function.

The operator \(\sigma_j^-\) is bosonic, i.e., operators on different sites commute, and it decreases \(d_{j,y} = \sigma_j^z\) by one, i.e.,
\beq
d_{j,y}\sigma_{j'}^- = \sigma_{j'}^-(d_{j,y} - \delta_{jj'})\punc.
\eeq
We seek a continuum operator with the corresponding properties. This can be constructed using the Hermitian operator \(\vartheta(x)\), defined in \refeq{eq:theta}, which is related to the conjugate field \(\Pi\) by \(\Pi(x) = -\pi \vartheta'(x)\) and obeys \([\vartheta(x),\vartheta(x')]=0\). From \refeq{eq:phivarthetacomm}, its commutator with \(\phi(x)\) is
\beq{}
[\phi(x),\vartheta(x')] = \frac{\ii}{\pi}\Theta_a(x-x')\punc,
\eeq
where, for \(0 \le x,x' < L_x\), \(\Theta_a\) obeys \(\Theta_a'(x)=\delta_a(x)\) and becomes a unit step function in the limit \(a \rightarrow 0\).

For any real \(m\), the operator \(V_m(x') = \ee^{2\pi\ii m \vartheta(x')}\) therefore shifts \(\phi(x)\) by \(-2m\Theta_a(x-x')\), i.e.,
\beq
\phi(x)V_m(x') = V_m(x') \left[\phi(x) - 2m\Theta_a(x-x')\right]\punc,
\eeq
and so
\beq
\partial_x\phi\,V_m(x') = V_m(x') \left[\partial_x\phi - 2m\delta_a(x-x')\right]\punc.
\eeq
Comparing with the first term in the bosonized (continuum) form of the operator \(d_{j[l],y}\), \refeq{eq:dyfinaloperator}, we therefore see that \(V_m(x_j)\) with \(m = \frac{1}{2}(-1)^{j+l}\) reduces the dimer number by one (spread over a region of size \(a\)).\footnote{The second term in \refeq{eq:dyfinaloperator} also makes a nonzero contribution over a region of size \(a\) near \(x_j\). This comes with a rapidly oscillating factor \((-1)^{j}\), and so makes no contribution to the long-distance correlation function.} We conclude that it is a continuum operator corresponding to the microscopic operator \(\sigma^-_{j[l]}\).

The monomer operator therefore corresponds to a ``magnetic vertex operator'' \cite{vonDelft1998,Nienhuis1987}, which inserts vortices in the height field, as has previously been stated in \refcites{Alet2006b,Fradkin2013,Allegra2015}. The fact that the magnetic charge \(m\) has opposite sign on even and odd sites reflects the opposite effective charge of monomers on the two sublattices \cite{Henley2010}. Since only charge-neutral products of vertex operators have nonzero expectation values, the appearance of \(\ee^{\pm\ii \pi \thetai}\) does not lead to ambiguities, even though (as explained in \refapp{bosonization}) \(\thetai\) is only defined up to an integer.

Note that \(V_{m,e}(x)=\ee^{2\pi\ii [ m \vartheta(x) + e \phi(x)]}\) is also a bosonic operator with the same commutation relations, as long as the ``electric charge'' \(e\) \cite{Nienhuis1987,Alet2006b} is an integer. In general, the microscopic operator can be expressed as a linear combination of \(V_{m,e}(x)\) with the same \(m\) but different \(e\). The long-distance form of the monomer distribution function will be dominated by \(V_m \equiv V_{m,0}\), since the others have correlation functions that decrease more rapidly with distance \cite{Nienhuis1987}.

Finally, we confirm that the proposed continuum operator gives the correct asymptotic form of the monomer distribution function \(G\sub{m}(\rv,\rv')\) for sites \(\rv\) and \(\rv'\) on opposite sublattices. If \(r_x+r_y\) is even and \(r_x'+r_y'\) is odd, then
\beq
G\sub{m}(\rv,\rv') \sim \left\langle V_{\frac{1}{2}}(\rv)V_{-\frac{1}{2}}(\rv')\right\rangle\punc.
\eeq
Because \(\vartheta\) is a free bosonic field, we have
\begin{equation}
\label{eq:vertexcorrelation1}
\langle \ee^{\ii \pi \vartheta(\rv)} \ee^{-\ii\pi\vartheta(\rv')}\rangle = \ee^{-\frac{\pi^2}{2}\left\langle\left[ \vartheta(\rv)-\vartheta(\rv')\right]^2\right\rangle}\punc.
\end{equation}
The Hamiltonian \(\ham\) in \refeq{eq:hamthetaphi}, when written in terms of \(\vartheta\) using \(\Pi(x) = -\pi\vartheta'(x)\), is symmetric between \(\vartheta\) and \(\phi\), and so their correlations are identical. We can therefore use \refeq{eq:asymptoticcorrelations} to give
\begin{equation}
\label{eq:vertexcorrelation}
G\sub{m}(\rv,\rv')\sim\langle \ee^{\ii \pi \vartheta(\rv)} \ee^{-\ii\pi\vartheta(\rv')}\rangle \sim \lvert \rv-\rv' \rvert^{-1/2}\punc,
\end{equation}
which is consistent with exact asymptotic results \cite{Hartwig1966,Wilkins2021}.

\section{Interacting double dimer model}
\label{interactingdoubledimermodel}

Up to this point, we have considered dimers that interact only through the close-packing constraint. In this section, we bosonize additional interactions between dimers and show how they modify the effective field theory. For simplicity, we restrict to the isotropic case, $\anis = 1$.

In particular, we consider the interacting double dimer model \cite{Wilkins2020}, consisting of two replicas of the close-packed dimer model on the square lattice, with coupling \(K\) between replicas 
and an aligning interaction \(J\) within each. The partition function is
\begin{equation}
Z\sub{4-GS} = \sum_{\substack{c\spr{1}\in\goC_0\\c\spr{2}\in\goC_0}} \ee^{-E\sub{4-GS}/T} \punc ,
\end{equation}
where \(c\spr{\alpha}\) is the configuration of replica \(\alpha\) and
\begin{equation}
\label{eq:configurationenergy}
E\sub{4-GS} = J\left[N_{\parallel}\spr{1}+N_{\parallel}\spr{2}\right]+KN\sub{o}
\end{equation}
is the total energy. Following the nomenclature of \refcite{Chen2009}, we refer to this as the 4-GS model, since it has four degenerate ground states when \(J<0\) and \(K<0\). In \refeq{eq:configurationenergy}, $N_{\parallel}\spr\alpha$ counts the number of parallel pairs of dimers in replica $\alpha$ \cite{Alet2005,Alet2006b},
\beq[eq:Nparallel]
N_\parallel\spr{\alpha} = \sum_{\langle l l'\rangle} \left[d_{l}\spr{\alpha} - \frac{1}{4}\right] \left[d_{l'}\spr{\alpha} - \frac{1}{4}\right]\punc,
\eeq
where the sum is over nearest-neighbor pairs of parallel bonds and \(d_l\spr{\alpha}\) is the dimer occupation number on bond \(l\) in replica \(\alpha\), while
\beq[eq:No]
N\sub{o} = \sum_{l} \left[d_{l}\spr{1} - \frac{1}{4}\right] \left[d_{l}\spr{2} - \frac{1}{4}\right]
\eeq
counts the number of dimers that overlap between the two replicas. In defining \(N_\parallel\) and \(N\sub{o}\), we choose to subtract from each \(d_l\spr{\alpha}\) its mean value \(\langle d_l\spr{\alpha}\rangle = \frac{1}{4}\) (by symmetry). This differs from the definition used in \refcite{Wilkins2020} by a constant shift in \(E\sub{4-GS}\), because \(\sum_l d_l\spr{\alpha} = \frac{1}{2}L_xL_y\) is fixed by the close-packing constraint.

\begin{figure}
\begin{center}
\includegraphics[width=\columnwidth]{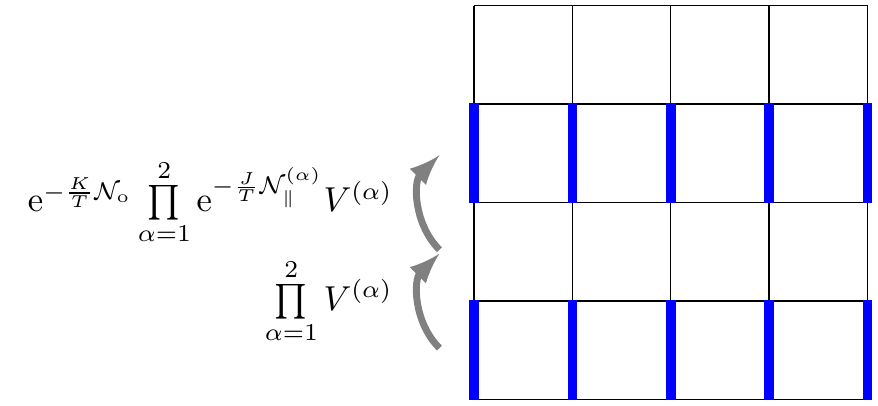}
\caption{Illustration of transfer matrix for the 1-GS double dimer model. Dimer--dimer interactions, \refeqand{eq:Nparallel}{eq:No}, are only applied on alternate rows of vertical bonds (blue), which we call type-1 bonds.}
\label{fig:1gs}
\end{center}
\end{figure}

Using the transfer-matrix formalism, it is simpler to consider a reduced-symmetry variant, the 1-GS model, illustrated in \reffig{fig:1gs}. In this model the aligning interactions and replica coupling are both applied only on alternate rows of vertical bonds, which we call type-1 bonds, giving a single ground state when \(J<0\) and \(K<0\). The configuration energy \(E\sub{1-GS}\) is given by \refeq{eq:configurationenergy} but with only type-1 bonds included in the sums in \refeqand{eq:Nparallel}{eq:No}. [Note that, summing over only type-1 bonds, \(\sum_l d_l\spr{\alpha}\) is not fixed. Subtracting \(\frac{1}{4}\) from each \(d_l\spr{\alpha}\) in \refeqand{eq:Nparallel}{eq:No} is equivalent to applying a potential on these bonds.]

In the following sections, we extend the transfer matrix to the 1-GS model. Because the interactions render the model nonintegrable \cite{Alet2006b}, we turn to perturbation theory. Instead of directly perturbing the eigenvalues and eigenvectors of $\ham$, we add perturbations to the effective field theory of \refeq{eq:sphi} by using bosonization. In doing so, we keep only terms of first order in \(J\) and \(K\), and also drop terms that cannot be relevant at the (Gaussian) RG fixed point corresponding to the free field theory.

Finally, in \refsec{4-gsmodel}, we use our results for the 1-GS model to infer the extension to the full 4-GS model. In \refcite{Wilkins2020}, we have shown that the latter exhibits a BKT phase transition between a standard Coulomb phase \cite{Henley2010} and a synchronized phase, and that the phase boundary passes through the noninteracting point $J=K=0$. We are therefore able to predict the shape of the phase boundary near this point by applying an RG analysis and perturbation theory in \(J\) and \(K\).

\subsection{Transfer-matrix perturbation theory}

Let $W\spr\alpha=\left[V\spr{\alpha}\right]^2=\ee^{-2\ham\spr\alpha}$ be the two-row transfer matrix for the noninteracting model in replica $\alpha$. The two-row transfer matrix for the 1-GS model is given by (see \reffig{fig:1gs})
\begin{equation}
\label{eq:v1gs2}
W\sub{1-GS} = \ee^{-\frac{K}{T}\mathcal{N}\sub o}\prod_{\alpha = 1}^{2}\ee^{-\frac{J}{T}\mathcal{N}_{\parallel}\spr\alpha}W\spr\alpha \punc ,
\end{equation}
where the operator
\begin{equation}
\label{eq:nparallel}
\mathcal{N}_\parallel\spr\alpha = \sum_{j=1}^{L_{x}} \left[d_{j,y}\spr\alpha - \frac{1}{4} \right] \left[d_{j+1,y}\spr\alpha - \frac{1}{4}\right]
\end{equation}
counts the number of parallel pairs of nearest-neighbor dimers in replica $\alpha$ on a row of vertical bonds, and
\begin{equation}
\label{eq:no}
\mathcal{N}\sub{o} = \sum_{j=1}^{L_{x}} \left[d_{j,y}\spr{1} - \frac{1}{4}\right]\left[d_{j,y}\spr{2} - \frac{1}{4}\right]
\end{equation}
counts the number of overlapping dimers on a row of vertical bonds.

To include these interactions in the transfer-matrix solution, we define the perturbed Hamiltonian $\ham\sub{1-GS}$ by
\begin{equation}
W\sub{1-GS} = \ee^{-2\ham\sub{1-GS}} \punc ,
\end{equation}
which, from \refeq{eq:v1gs2}, is given by
\begin{equation}
\label{eq:ham1gs}
\ham\sub{1-GS} = -\frac{1}{2}\log\left(\ee^{-\frac{K}{T}\mathcal{N}\sub o}\ee^{-\frac{J}{T}\left[\mathcal{N}_{\parallel}\spr1+\mathcal{N}_{\parallel}\spr2\right]}\ee^{-2\left[\ham\spr1+\ham\spr2\right]}\right) \punc .
\end{equation}
The operators $\mathcal{N}_{\parallel}\spr\alpha$ and $\mathcal{N}\sub{o}$ commute with each other, but not with $\ham\spr\alpha$, and so the Baker--Campbell--Hausdorff formula, \refeq{eq:BCH}, is required to simplify \refeq{eq:ham1gs}. To first order in the couplings \(J\) and \(K\) but all orders in \(\ham\spr\alpha\), this can be written as
\begin{multline}
\label{eq:hamBCH1}
\ham\sub{1-GS} = \sum_{\alpha=1}^{2}\left(\ham\spr\alpha + \frac{1}{2}\frac{J}{T}\mathcal{N}_{\parallel}\spr\alpha\right) + \frac{1}{2}\frac{K}{T}\mathcal{N}\sub{o}\\
+ \sum_{n=1}^\infty c_n\left\{\frac{K}{T}\mathcal{L}^n\mathcal{N}\sub{o} + \frac{J}{T}\mathcal{L}^n\left[\mathcal{N}_{\parallel}\spr1+\mathcal{N}_{\parallel}\spr2\right]\right\}\punc ,
\end{multline}
for some coefficients \(c_n\) \cite{Moodie2020}, where \(\mathcal{L}\) is a superoperator defined by
\beq
\mathcal{L}X = \left[\ham\spr1+\ham\spr2, X\right]\punc.
\eeq

In fact, we need only retain terms on the first line of \refeq{eq:hamBCH1}. To see this, note that \refeq{eq:ham2} gives $[\ham,\zeta_{k}]=-\epsilon(k)\zeta_{k}$ (omitting replica indices). Since $\mathcal{N}_{\parallel}\spr\alpha$ and $\mathcal{N}\sub{o}$ are polynomials in \(\zeta_k\), each action of \(\mathcal{L}\) gives an additional factor of \(\epsilon(k)\), producing terms that are less important in the long-wavelength limit. We therefore drop all commutators in \refeq{eq:hamBCH1}, leaving simply
\beq[eq:hamBCH]
\ham\sub{1-GS} \approx \sum_{\alpha=1}^{2}\left(\ham\spr\alpha + \frac{1}{2}\frac{J}{T}\mathcal{N}_{\parallel}\spr\alpha\right) + \frac{1}{2}\frac{K}{T}\mathcal{N}\sub{o}\punc.
\eeq

In the following sections we linearize and then bosonize $\mathcal{N}_{\parallel}\spr\alpha$ and $\mathcal{N}\sub{o}$, giving expressions in terms of the bosonic operators \(\phi\) and \(\Pi\). We then use these results, along with \refeq{eq:hamBCH}, to find the corresponding perturbation to the 1-GS model, to first order in the couplings $J$ and $K$.

\subsection{Aligning interactions}
\label{aligninginteractionsnparallel}

The calculation of $\mathcal{N}_{\parallel}$ (we drop replica indices in this section) in the free theory follows the procedure in \refsec{dimeroccupationnumbers} for the dimer occupation numbers. However, because this operator is quartic, rather than quadratic, in fermions, the algebra is more involved.

We first write \refeq{eq:nparallel} in terms of fermion operators and then normal order using \refeq{eq:no4}, giving
\begin{widetext}
\begin{equation}
\label{eq:nparallelno}
\mathcal{N}_{\parallel} = \frac{L_{x}}{16} + \sum_{j=1}^{L_{x}}\bigg[\no{C_{j}^{\dagger}C_{j}C_{j+1}^{\dagger}C_{j+1}} +\frac{1}{4}\no{C_{j}^{\dagger}C_{j+1}^{\dagger} - C_{j}C_{j+1}}\bigg] \punc .
\end{equation}
(We define normal ordering using the ground state of the noninteracting dimer model, and so expectation values are calculated in the latter.)

To bosonize this, we use \refeq{eq:clr} with \(l=0\) to express \(C_j\) in terms of the continuum fermion operators \(R\) and \(L\). For the quartic term, this gives
\begin{multline}
\label{eq:4fermionterm}
\no{C_{j}^{\dagger}C_{j}C_{j+1}^{\dagger}C_{j+1}} \approx -\no{\left[\hc{R}(x)R(x)+\hc{L}(x)L(x)\right]\left[\hc{R}(x')R(x')+\hc{L}(x')L(x')\right]} \\
+\no{\left[\hc{R}(x)L(x)+\hc{L}(x)R(x)\right] \left[\hc{R}(x')L(x')+\hc{L}(x')R(x')\right]}\punc,
\end{multline}
\end{widetext}
where \(x=x_j\) and \(x'=x_{j+1}\) are the positions of two adjacent sites. In writing \refeq{eq:4fermionterm}, we have dropped terms that alternate in sign for even and odd \(j\), which give negligible contributions when summed over \(j\).

Since we are retaining only the most RG-relevant terms, i.e., those without derivatives, we can replace \(x' \rightarrow x\), effectively keeping only the zeroth-order term in a Taylor expansion in \(x'-x\). All terms that involve repeated operators, such as \(R(x)R(x')\), therefore vanish and we are left with
\beq
\no{C_{j}^{\dagger}C_{j}C_{j+1}^{\dagger}C_{j+1}} \approx 
-4\no{\hc{R}(x)R(x)}\no{\hc{L}(x)L(x)}\punc,
\eeq
where we have used the fact that all fermion operators anticommute within normal ordering (see \refapp{fermions}). Using \refeqand{eq:dxphi}{eq:Pi} in \reftab{tab:bosonizationdictionary}, we find
\begin{equation}
\label{eq:-2LLRR}
-4\no{R^{\dagger}(x)R(x)}\no{L^{\dagger}(x)L(x)} = \frac{1}{\pi^2}\Pi^{2} - (\partial_{x}\phi)^{2}\punc,
\end{equation}
omitting a boundary term involving \(p\) that makes a negligible contribution in the thermodynamic limit.

Following the same steps, the quadratic terms in \refeq{eq:nparallelno} give
\beq
\no{C_{j}^{\dagger}C_{j+1}^{\dagger} - C_{j}C_{j+1}} \approx 2\ii (-1)^j \no{L^\dagger(x)R(x) - R^\dagger(x)L(x)}\punc,
\eeq
and hence make no contribution to this order.\footnote{If one does not subtract \(\frac{1}{4}\) from each \(d_l\spr{\alpha}\) in \refeq{eq:Nparallel}, the quadratic terms make an additional contribution \(\frac{1}{2\pi a}\sin\left(2\pi\phi\right)\) \cite{WilkinsThesis}. According to the logic following \refeq{eq:No}, this cannot have any effect in the 4-GS model.}

The result is therefore
\begin{equation}
\label{eq:nparthetaphi}
\mathcal{N}_{\parallel} = \frac{L_{x}}{16} + \int_{0}^{L_{x}} \dd x \left[\bno{\frac{1}{\pi^2}\Pi^{2} - (\dx\phi)^{2}}\right] \punc ,
\end{equation}
where we have normal-ordered the bosonic operators to match those in the Hamiltonian, \refeq{eq:hamthetaphi}. Because it is quadratic in boson operators, this simply adds a constant, but using \refeq{eq:bnophi2} one can show that this constant is in fact zero.

This result, when combined with \refeq{eq:hamBCH}, shows that the effect of aligning interactions is to modify the relative coefficients of the two terms in the unperturbed Hamiltonian, \refeq{eq:hamthetaphi}. This is in qualitative agreement with \refcite{Papanikolaou2007}, where $\mathcal{N}_{\parallel}$ is found by substituting \refeq{eq:dyfinaloperator} into the continuum version of \refeq{eq:nparallel} and using an operator product expansion. The results are not in quantitative agreement, however, which we believe is due to inconsistent treatment of short-distance regularization in \refcite{Papanikolaou2007}.\footnote{Specifically, in \refcite{Papanikolaou2007} the short-distance cutoff \(\mu\) is set equal to \(1\) (in units of the lattice spacing) rather than \(1/\pi\), as required by their expression for the dimer operators in terms of \(\phi\) (see Footnote~\ref{footnote:cutoff}).}

\subsection{Replica coupling}
\label{replicacouplingno}

In contrast to $\mathcal{N}_{\parallel}$, one can calculate $\mathcal{N}\sub{o}$ directly by substituting the bosonized expression for \(d_y\spr{\alpha}\), \refeq{eq:dyfinaloperator}, into the continuum version of \refeq{eq:no},
\begin{equation}
\mathcal{N}\sub{o} = \int_{0}^{L_{x}} \dd x \, \left[d_{y}\spr{1} - \frac{1}{4}\right]\left[d_{y}\spr{2}-\frac{1}{4}\right] \punc .
\end{equation}
This is because operators with different replica indices commute and hence, provided $d_{y}\spr{1}$ and $d_{y}\spr{2}$ are each fermion normal ordered, the product $d_{y}\spr{1}d_{y}\spr{2}$ is also normal ordered. An analogous calculation for the Hubbard model, with spin playing the role of replica index, can be found in \refcite{Giamarchi2004}.

Making this substitution and dropping terms with a factor of \((-1)^j\) as in \refeq{eq:4fermionterm}, one obtains
\begin{multline}
\label{eq:nothetaphi}
\mathcal{N}\sub{o} = \int_{0}^{L_{x}} \dd x \, \bigg[\left(\partial_{x}\phi\spr{1}\right)\left(\partial_{x}\phi\spr{2}\right) \\
+\frac{1}{(\pi a)^{2}}\sin{\left(2\pi\phi\spr{1}\right)}\sin{\left(2\pi\phi\spr{2}\right)}\bigg]\punc.
\end{multline}

\subsection{1-GS model}
\label{1-gsmodel}

To find the bosonized Hamiltonian for the 1-GS model, we substitute the expressions for the aligning interactions \(\mathcal{N}_{\parallel}\), \refeq{eq:nparthetaphi}, and the replica coupling \(\mathcal{N}\sub{o}\), \refeq{eq:nothetaphi} into \refeq{eq:hamBCH}. This gives, up to additive constants,
\begin{widetext}
\begin{multline}
\label{eq:ham1gs2}
\ham\sub{1-GS} = 
\int_{0}^{L_{x}} \dd x \,\Bigg[\sum_{\alpha = 1}^{2}\frac{1}{2}\bno{{\kappa_{\text{1-GS},x}}\left(\partial_{x}\phi\spr\alpha\right)^{2} + \kappa_{\text{1-GS},y}^{-1}\left(\Pi\spr\alpha\right)^{2}}\\
+ {\lambda\sub{1-GS}}\left(\partial_{x}\phi\spr{1}\right)\left(\partial_{x}\phi\spr{2}\right)
+ V\sub{1-GS}\sin{\left(2\pi\phi\spr{1}\right)}\sin{\left(2\pi\phi\spr{2}\right)}\Bigg]\punc,
\end{multline}
where
\begin{align}
\label{eq:kappax}
\kappa_{\text{1-GS},x} &= \pi\left(1 - \frac{1}{\pi}\frac{J}{T} \right) \\
\label{eq:kappay}
\kappa_{\text{1-GS},y} &= \pi\left(1 + \frac{1}{\pi}\frac{J}{T} \right)^{-1}\approx \pi\left(1 - \frac{1}{\pi}\frac{J}{T}\right) \\
\label{eq:lambdax}
\lambda\sub{1-GS} &= \frac{1}{2}\frac{K}{T} \\
V\sub{1-GS} &= \frac{1}{2}\frac{K}{T}\frac{1}{(\pi a)^{2}} \punc . 
\end{align}
The approximations made in the derivation of \refeq{eq:ham1gs2} can only modify these coefficients at higher order in $J/T$ and $K/T$ and give other terms in the Hamiltonian that are less relevant under the RG \cite{Papanikolaou2007}.

The corresponding action is given by \(S\sub{1-GS} = S\spr{1}+S\spr{2}+\delta S\sub{1-GS}\), where \(S\spr{\alpha}\) is the unperturbed action for replica \(\alpha\) [\refeq{eq:sphi} with \(\anis = 1\)] and
\begin{equation}
\label{eq:1-gsaction}
\delta S\sub{1-GS} = \int \dd^2 \rv\, \left[\frac{1}{2}\delta\kappa\sub{1-GS}\left(\left\lvert\del\phi\spr{1}\right\rvert^{2}+ \left\lvert\del\phi\spr{2}\right\rvert^{2} \right)
+ \lambda\sub{1-GS}\left(\partial_{x}\phi\spr{1}\right)\left(\partial_{x}\phi\spr{2}\right)
+ V\sub{1-GS}\sin{\left(2\pi\phi\spr{1}\right)}\sin{\left(2\pi\phi\spr{2}\right)}\right]\punc,
\end{equation}
\end{widetext}
where \(\delta\kappa\sub{1-GS} = -J/T\). The first term is a modification of the stiffness \(\kappa\) due to $J$, while the second is a quadratic coupling between replicas due to \(K\). (Note that, even though both types of interaction are anisotropic, anisotropy in the action is only generated by $K$ at this order.) The final term is the most RG-relevant periodic function of \(\phi\spr{1,2}\) that is consistent with symmetry constraints, and is allowed only because the 1-GS model breaks symmetry under fourfold rotations \(\mathbb{R}\).

\subsection{4-GS model}
\label{4-gsmodel}

The perturbation to the action, \(\delta S\sub{1-GS}\), gives the effect of interactions applied only on the type-\(1\) bonds. In this section, we use the symmetry properties of \(\phi\), given in \refeqand{eq:Tphi}{eq:Rphi}, to infer the action for the 4-GS model, where the interactions are applied with equal strength on all bonds.

\begin{figure}
\begin{center}
\includegraphics[width=0.7\columnwidth]{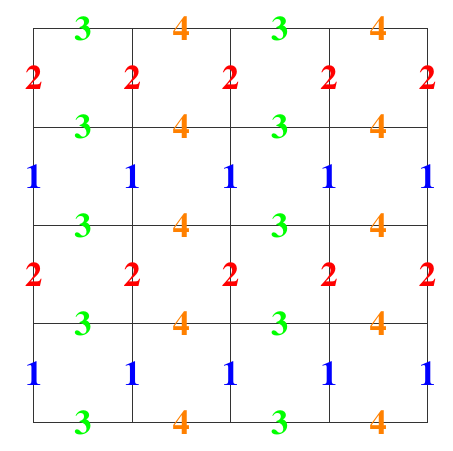}
\caption{Definition of bond types $i\in\{1,2,3,4\}$ in the square lattice. In the 1-GS model (see \reffig{fig:1gs}), the couplings \(J\) and \(K\) are applied only on type-1 bonds. In the 4-GS model, they are applied on all bond types with equal strength.}
\label{fig:bondlabels}
\end{center}
\end{figure}

To do so, we decompose the lattice into four bond types $i\in\{1,2,3,4\}$, defined in \reffig{fig:bondlabels}. The effect of interactions on bonds of type \(2\) to \(4\) can be found by applying \(\mathbb{T}_y\) and \(\mathbb{R}\) to \(\delta S\sub{1-GS}\), which has interactions only on bonds of type \(1\). At first order in perturbation theory, the change in the effective action is linear in the perturbation, and so the total perturbation to the action for the 4-GS model is simply given by the sum of the contributions from the four types of bond.

To find the contribution from bonds of type 2, we apply \(\mathbb{T}_y\), which maps \(\phi\spr{\alpha}\rightarrow -\phi\spr{\alpha}\) and hence leaves \(\delta S\sub{1-GS}\) unchanged. For type-4 bonds, we apply \(\mathbb{R}\), mapping \(\sin\left(2\pi\phi\spr{\alpha}\right)\rightarrow\cos\left(2\pi\phi\spr{\alpha}\right)\) and \((\partial_x,\partial_y) \rightarrow (-\partial_y,\partial_x)\). Finally, for type-3 bonds we apply \(\mathbb{T}_y\) followed by \(\mathbb{R}\), giving the same as for type-4. The total action is therefore \(S\sub{4-GS} = S\spr{1}+S\spr{2}+\delta S\sub{4-GS}\) with
\begin{equation}
\label{eq:4-gsaction}
\begin{aligned}
\delta S\sub{4-GS} = \int \dd^2 \rv\, \Bigg\{&\frac{1}{2}\delta\kappa\left(\left\lvert\del\phi\spr{1}\right\rvert^{2}+ \left\lvert\del\phi\spr{2}\right\rvert^{2} \right) \\
&+ \lambda\left(\del\phi\spr{1}\right)\cdot\left(\del\phi\spr{2}\right) \\
&+ V\spr{-}\cos{\left[2\pi\left(\phi\spr{1}-\phi\spr{2}\right)\right]}\Bigg\}\punc,
\end{aligned}
\end{equation}
where
\begin{align}
\delta\kappa &= 4\delta\kappa\sub{1-GS} = -\frac{4J}{T}\\
\lambda &= 2\lambda\sub{1-GS} = \frac{K}{T}\\
V\spr{-} &= 2V\sub{1-GS} = \frac{K}{T}\frac{1}{(\pi a)^{2}}\punc.
\end{align}
One can diagonalize the quadratic terms by defining \cite{Wilkins2020}
\beq[eq:phipm]
\phi\spr\pm = \phi\spr1 \pm \phi\spr2
\punc ,
\eeq
in terms of which
\begin{multline}
\label{eq:4-gsactionpm}
S\sub{4-GS} = \int \dd^2 \rv\, \Bigg\{\frac{1}{2}\kappa_+\left\lvert\del\phi\spr{+}\right\rvert^{2}+ \frac{1}{2}\kappa_-\left\lvert\del\phi\spr{-}\right\rvert^{2} \\
+ V\spr{-}\cos{\left(2\pi\phi\spr{-}\right)}\Bigg\}
\punc,
\end{multline}
with
\begin{align}
\label{eq:kappaxpm}
\kappa_{\pm} &= \frac{\pi}{2} + \frac{1}{2}(\delta\kappa \pm \lambda)\\
&= \frac{\pi}{2} -\frac{2J}{T} \pm \frac{K}{2T}\punc.
\end{align}
This form for \(S\sub{4-GS}\), including the values of \(\kappa_\pm\) at \(J=K=0\), was deduced in \refcite{Wilkins2020}.\footnote{A term \(\cos\left(4\pi \phi\spr+\right)\) is also allowed by symmetry \cite{Wilkins2020}, but is not important near \(J=K=0\) and does not appear at this order in perturbation theory.} Here, we have also determined how the parameters $\kappa_{\pm}$ and $V\spr{-}$ depend on the microscopic couplings $J$ and $K$.

\begin{figure}
\begin{center}
\includegraphics[width=\columnwidth]{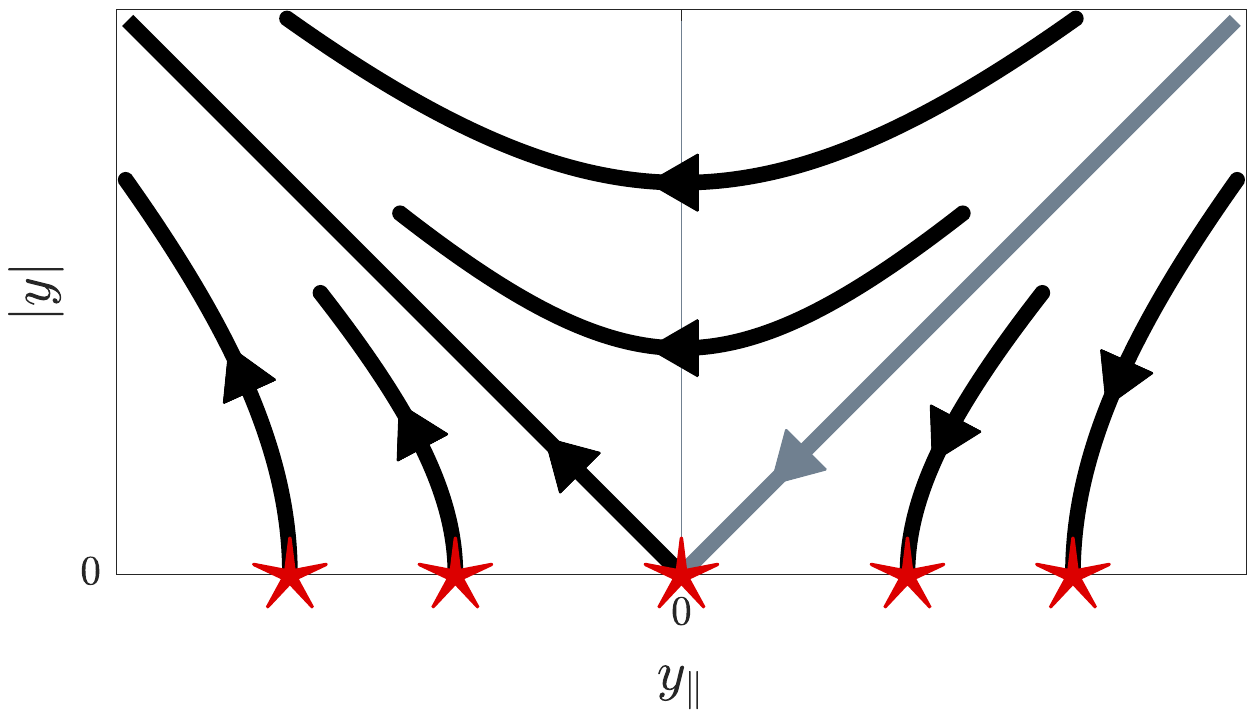}
\caption{RG flows of the sine-Gordon action, \refeq{eq:sggen}, with $y_{\parallel}=2(\mathcal{K}-1)$ and $y=g/\pi$. There is a line of fixed points (red stars) along $\lvert y \rvert = 0$, which are stable (unstable) for $y_{\parallel}>0$ ($y_{\parallel}<0$). The trajectories are hyperbolas, and the separatrix $\lvert y \rvert = y_{\parallel}$ (gray) divides regions where $\lvert y\rvert$ flows to zero and infinity.}
\label{fig:RGflow}
\end{center}
\end{figure}

In order to determine the critical properties of $S\sub{4-GS}$, we appeal to an RG analysis, including the leading-order effect of \(V\spr-\) on the flows. For the general sine-Gordon action \cite{Giamarchi2004}
\begin{equation}
\label{eq:sggen}
S\sub{SG}[\phi] = \int \dd^2\rv \bigg[ \frac{1}{2} \frac{\pi}{2\mathcal{K}}\lvert\del\phi\rvert^{2} + \frac{1}{2}\frac{g}{(\pi a)^{2}}\cos(2\pi\phi)\bigg]\punc ,
\end{equation}
one can show\footnote{To be explicit, our \refeq{eq:sggen} is the action corresponding to the Hamiltonian in Eq.~(2.106) of \refcite{Giamarchi2004}, with the replacement \(\phi \rightarrow \frac{\pi}{\sqrt{2}}\phi\), and the flow equations are Eq.~(2.140).} that the RG flow equations for $\lvert g \rvert \ll 1$ are given by
\begin{align}
\label{eq:flows}
\frac{\dd y_{\parallel}}{\dd\ell} &= -y^{2} & \frac{\dd y}{\dd\ell} &= -y_{\parallel}y \punc ,
\end{align}
where $y_{\parallel} = 2(\mathcal{K}-1)$, $y=g/\pi$, and \(\ell\) is a scale parameter. The RG flows described by these equations are illustrated in \reffig{fig:RGflow}. In particular, since $y \,\dd y=y_{\parallel}\,\dd y_{\parallel}$, we have
\begin{equation}
\label{eq:hyperbola}
y_{\parallel}^{2} - y^{2} = \text{const} \punc ,
\end{equation}
i.e., the trajectories are hyperbolas. There is a BKT phase transition along the separatrix \(\lvert y\rvert = y_{\parallel}\): For \(y_\parallel > \lvert y\rvert\), $y$ flows to zero, meaning that the cosine term renormalizes to zero at long distances leaving a free Gaussian theory. For \(y_\parallel < \lvert y\rvert\), $y$ flows to \(\pm\infty\) giving a dominant cosine term that locks \(\phi\) to discrete values.

In the case of the 4-GS model, the field \(\phi\spr{-}\) is governed by a sine-Gordon action with
\begin{align}
y_\parallel &= 2\left(\frac{\pi}{2\kappa_{-}}-1\right) \approx \frac{2}{\pi}\left(4\frac{J}{T}+\frac{K}{T}\right)\\
y&=\frac{2}{\pi}\frac{K}{T}\punc.
\end{align}
The separatrix at \(\lvert y\rvert = y_{\parallel}\) corresponds to a phase boundary at
\begin{equation}
\label{eq:phaseboundary}
\left(\dfrac{J}{T}\right)\sub{c} = \begin{cases}
0 & \text{for \(\dfrac{K}{T}\geq 0\)} \\[6pt]
-\dfrac{1}{2}\dfrac{K}{T} & \text{\phantom{for} \(\dfrac{K}{T}\leq 0\),}
\end{cases}    
\end{equation}
with \(\phi\spr{-}\) locked to discrete values for \(\left(\frac{J}{T}\right) < \left(\frac{J}{T}\right)\sub{c}\). As shown in \refcite{Wilkins2020}, locking of the relative field \(\phi\spr-\) corresponds to either the synchronized or antisynchronized phase of the double dimer model, depending on the sign of \(K\).

The result in \refeq{eq:phaseboundary} therefore constitutes a quantitative prediction for the asymptotic behavior of the phase boundary near the noninteracting point \(J=K=0\). This prediction is indeed consistent with our numerical determination of the phase boundary in Fig.~3 of \refcite{Wilkins2020}, which used a Monte Carlo worm algorithm. In particular, the data point closest to the origin is $(K/T,J/T)=(-0.097(9),0.05)$, in agreement with \refeq{eq:phaseboundary} for \(K/T \le 0\), while the phase boundary runs along the line $J/T=0$ for $K/T \ge 0$, as was previously conjectured \cite{Raghavan1997}.

\section{Conclusions}
\label{conclusions}

Starting from the transfer-matrix solution of the classical dimer model on the square lattice, we have used bosonization to derive a continuum field theory to describe its long-wavelength properties. The result, expressed by the action of \refeq{eq:sphi} and the relationships \refeqand{eq:dyfinalfield}{eq:dxfinal}, is a free, massless real field theory in 2D and is consistent with the ``height theory'' that is well established as the continuum description of the dimer model \cite{Fradkin2013}.

In contrast to previous works, which justify the continuum description based on general properties such as symmetry, our constructive derivation gives exact values for the coefficients in the action (including in the anisotropic case) and those relating the continuum fields to microscopic observables. Other properties of the field theory also arise naturally from our derivation, without needing to be added ``by hand'', such as the fact that the height is defined only up to an integer and that it has boundary discontinuities related to the flux.

In addition, our approach allows us to apply perturbation theory to treat weak interactions added to the noninteracting dimer model. In the fermionic language, the Coulomb phase of the original dimer model is a noninteracting Fermi gas, while adding interactions gives a Luttinger liquid. In the double dimer model, this has two components in the unsynchronized Coulomb phase but a single component in the synchronized phase. Our derivation determines the leading-order perturbation to the action due to the microscopic interactions, allowing a quantitative prediction of the shape of the phase boundary near the noninteracting point.

The long-wavelength theory derived here can be extended to include higher-order terms and hence determine corrections to the asymptotic forms of correlation functions. These include corrections to the relations between dimer observables and operators in the field theory \cite{WilkinsThesis} as well as corrections to the action due to band curvature \cite{Teber2007,Pereira2007}.

The bosonization approach could straightforwardly be applied to dimer models on other planar bipartite lattices, such as the honeycomb lattice, and to other exactly-solved models whose transfer matrix has a free-fermion form. In fact, for any such model, the long-wavelength Hamiltonian will take the form of \refeq{eq:firstorder}, with \(w\) given by the Fermi speed, and hence the action will be identical to \refeq{eq:sphi}. (This follows from the fact that any free-fermion model has Luttinger parameter \(K=1\) \cite{Sachdev2011}.) Note, however, that relationships between the microscopic degrees of freedom and the continuum fields, such as \refeqand{eq:dyfinalfield}{eq:dxfinal}, will not be the same, including in the honeycomb-lattice dimer model \cite{Fradkin2004}.

The approach may also be applicable to the dimer model on the triangular lattice \cite{Fendley2002}, which is not bipartite, but can be treated as a square lattice with added diagonal links. These give operators that create or annihilate pairs of monomers, and hence a term of the form \(\cos(2\vartheta)\) in the field theory \cite{Trousselet2007}.

\acknowledgments

This work was supported by EPSRC Grant No.\ EP/T021691/1.

\appendix

\section{Baker--Campbell--Hausdorff formula}
\label{bakercampbellhausdorffformula}

The first few terms in the Baker--Campbell--Hausdorff formula are \cite{Rossmann2006}
\begin{equation}
\begin{split}
\label{eq:BCH}
\log{\left(\ee^{A}\ee^{B}\right)} = {} &A + B + \frac{1}{2}[A,B] + {} \\ &\frac{1}{12}\left([A,[A,B]] + [B,[B,A]\right) + \dotsc \punc ,
\end{split}
\end{equation}
where the omitted terms involve three or more nested commutators.

In the case $[A, [A,B]] = [B, [A,B]] = 0$, \refeq{eq:BCH} implies
\begin{equation}
\label{eq:corollary1}
\ee^{A}\ee^{B} = \ee^{A+B}\ee^{[A,B]/2} \punc ,
\end{equation}
and
\begin{equation}
\label{eq:corollary2}
\ee^{A}\ee^{B} = \ee^{B}\ee^{A}\ee^{[A,B]} \punc .
\end{equation}

\section{Normal ordering}
\label{normalordering}

In this appendix we define fermion and boson normal ordering, denoted by $\no{\quad}$ and \(\bno{\quad}\) respectively. Boson and fermion normal ordering are not in general equivalent \cite{Rao2001}.

\subsection{Fermions}
\label{fermions}

For fermions, normal ordering is required when relating the microscopic lattice model to the long-wavelength continuum theory. It is defined with respect to a particular choice of a reference many-body state \(\lvert 0 \rangle\).

In both parity sectors of the microscopic model, we take this state such that
\beq[eq:refstatezetak]
\begin{aligned}
&\zeta_{k}\lvert 0 \rangle = 0 \qquad &\text{for} \qquad 0 &< k < \pi \\
&\zeta^{\dagger}_{k}\lvert 0 \rangle = 0 \qquad &\text{\phantom{for}}\qquad -\pi &\le k \le 0
\punc,
\end{aligned}
\eeq
so that all single-fermion states with \(\epsilon(k) < 0\) are occupied and all those with \(\epsilon(k) > 0\) are empty. In the sector with odd parity, \(p=1\), there are also single-fermion states at \(k=0\) and \(k=-\pi\), for which \(\epsilon(k) = 0\); in \refeq{eq:refstatezetak}, both are occupied and so, using \refeq{eq:fluxzeta}, \(\Phi_y = +1\).

The notation $\no{\quad}$ then means to anticommute all $\zeta_{k}^{\dagger}$ with $-\pi \le k \le 0$ and all $\zeta_{k}$ with $0 < k < \pi$ to the right of all other creation and annihilation operators. For example, with $-\pi \le k \le 0$,
\begin{equation}
\no{\zeta_{k}^{\dagger}\zeta_{k}} = -\zeta_{k}\zeta_{k}^{\dagger} \punc .
\end{equation}

Similar definitions apply for for the right- and left-moving fermions of the continuum theory, \(\Psi_{k+} = R_k\) and \(\Psi_{k-}=L_k\). In this case, the reference state \(\lvert 0 \rangle\) has all single-fermion states with \(k \le 0\) occupied and all others empty, and $\no{\quad}$ is defined by anticommuting all $R_{k}^{\dagger}$ and \(L_k^\dagger\) with $k\le 0$ and all $R_{k}$ and \(L_k\) with $k>0$ to the right of all other \(R_k^{(\dagger)}\) and \(L_k^{(\dagger)}\) operators.

In the case of two fermion operators, the above definitions of normal ordering are equivalent to
\begin{equation}
\label{eq:no2}
AB = \no{AB} + \langle 0 \rvert AB \lvert 0 \rangle \punc .
\end{equation}
For four fermion operators one requires the Wick expansion \cite{Fetter2003,Rozhkov2005,Rozhkov2006}
\begin{equation}
\label{eq:no4}
\begin{split}
AB&CD = \no{ABCD} + {} \\ &\vev{AB}\no{CD} - \vev{AC}\no{BD} + \vev{AD} \no{BC} + {} \\ &\vev{BC}\no{AD} - \vev{BD}\no{AC} + \vev{CD}\no{AB} + {} \\ &\vev{AB}\vev{CD} - \vev{AC}\vev{BD} + \vev{AD}\vev{BC} \punc ,
\end{split}
\end{equation}
where $\langle \cdot\rangle_{0} \equiv \langle 0 \lvert \cdot \rvert 0 \rangle$.

Note that any two fermion operators anticommute when they are within normal ordering, e.g.,
\begin{equation}
\label{eq:swap}
\no{ABCD} = -\no{BACD} \punc .
\end{equation}

\subsection{Bosons}
\label{bosons}

For boson operators, $\bno{\quad}$ means to commute all annihilation operators $b_{q\eta}$ to the right of all creation operators $b_{q\eta}^{\dagger}$ of the same species, e.g.,
\begin{equation}
\bno{b_{q\eta}b_{q'\eta}^{\dagger}} = b_{q'\eta}^{\dagger}b_{q\eta} \punc .
\end{equation}

We include an explicit short-distance regularization \(a\) in the bosonized theory, and expressions where bosonic operators are not in normal order typically depend on \(a\), while normal-ordered expressions do not.

\section{Bosonization}
\label{bosonization}

This appendix provides a brief overview of (abelian) bosonization. The main results are summarized in \reftab{tab:bosonizationdictionary}. For an application to the XXZ spin chain using a similar approach, see \refcite{WilkinsThesis}.

\subsection{Bosonization identity}
\label{bosonizationidentity}

In this section we state the bosonization identity, which is an operator identity in Fock space, as well as defining the fermion and boson fields that appear within it. We follow \refcite{vonDelft1998} but with some changes to definitions of the fields; see \refcites{Giamarchi2004,Sachdev2011} for other useful resources.

We start from right- and left-moving fermions as introduced in \refsec{linearizedtheory}. In this appendix, we use the notation \(\Psi_+ \equiv R\) and \(\Psi_- \equiv L\) for the right- and left-moving fermions, respectively. They obey
\begin{align}
\label{eq:fourierseries2}
&\Psi_\eta(x) = \frac{1}{\sqrt{L_{x}}}\sum_{k}\ee^{\ii \eta kx}\Psi_{k\eta} \\
&\Psi_{k\eta} = \frac{1}{\sqrt{L_{x}}}\int_{0}^{L_{x}} \dd x \, \ee^{-\ii \eta kx} \Psi_\eta(x) \punc ,
\end{align}
for \(\eta = \pm\), where $L_{x}$ is the number of lattice sites. Here and in the following, the momenta included in the sum over \(k\) are the same as in \refeq{eq:fs},
\begin{equation}
k \in \frac{2\pi}{L_{x}}\left[ \mathbb{Z} + \frac{1}{2}(1-p)\right] \punc ,
\end{equation}
where the parameter \(p = 0\) or \(1\).\footnote{In the notation of \refcite{vonDelft1998}, this corresponds to \(\delta\sub{b}=1-p\).} This implies boundary conditions on \(\Psi_\eta(x)\) of
\beq[eq:psibc]
\Psi_\eta(x+L_x) = -(-1)^p\Psi_\eta(x)\punc.
\eeq
The commutation relations are
\beq[eq:Psicomm]
\begin{aligned}
\{\Psi_{k\eta},\Psi_{k'\eta'}^\dagger\} &= \delta_{\eta\eta'}\delta_{kk'}&
\{\Psi_{k\eta},\Psi_{k'\eta'}\} &= 0\\
\{\Psi_{\eta}(x),\Psi_{\eta'}^\dagger(x')\} &= \delta_{\eta\eta'}\delta(x-x')&
\{\Psi_{\eta}(x),\Psi_{\eta'}(x')\} &= 0
\end{aligned}
\eeq
for \(0 \le x,x' < L_x\).

For each \(\eta\), we define the fermion number operator
\begin{equation}
\label{eq:Neta}
N_\eta = \sum_{k}\no{\Psi_{k\eta}^{\dagger}\Psi_{k\eta}} = \int_0^{L_x} \dd x\, \no{\Psi_\eta^\dagger(x)\Psi_\eta(x)}
\punc ,
\end{equation}
which counts the number of fermions of type \(\eta\) relative to the reference state \(\lvert 0 \rangle\). Normal ordering, denoted \(\no{\quad}\), is defined in \refapp{fermions}; as there, we take the reference state to obey
\beq[eq:refstate]
\begin{aligned}
& \Psi_{k\eta}\lvert 0 \rangle = 0 \qquad \text{for} \qquad k > 0 \punc , \\
& \Psi^{\dagger}_{k\eta}\lvert 0 \rangle = 0 \qquad \text{\phantom{for}} \qquad k \le 0 \punc ,
\end{aligned}
\eeq
and hence to be the ground state of the Hamiltonian, \refeq{eq:firstorder}. (If \(p=1\), then \(k=0\) is included and \(\lvert 0 \rangle\) is one of four degenerate ground states.)

One can construct bosonic operators from bilinears in the fermionic operators. First, we define creation operators
\begin{equation}
b_{q\eta}^{\dagger} = \frac{\ii}{\sqrt{n_{q}}}\sum_{k}\Psi_{k+q,\eta}^{\dagger}\Psi_{k,\eta}
\end{equation}
for $q={2\pi n_{q}}/{L_{x}}$ with $n_{q} \in \mathbb{Z}^{+}$, and corresponding annihilation operators \(b_{q\eta} = \big(b_{q\eta}^\dagger\big)^\dagger\). They obey \cite{vonDelft1998}
\begin{align}
[ b_{q \eta},b_{q' \eta'}^{\dagger}] &= \delta_{\eta\eta'}\delta_{qq'} & [b_{q\eta},b_{q'\eta'}]&=0
\punc,
\end{align}
and commute with the fermion number operators \(N_\eta\), since they create particle--hole pairs. Note also that, since \(b_{q\eta}\) contains only operators \(\Psi_{k\eta}^\dagger\Psi_{k'\eta}\) with \(k' > k\), it obeys \(b_{q\eta}\lvert 0 \rangle = 0\).

We also define real-space operators
\begin{equation}
\label{eq:chiralladderoperators}
\varphi_{\eta}^{\dagger}(x) = \frac{\eta}{2\pi}\sum_{q>0}\frac{1}{\sqrt{n_{q}}}\ee^{-\ii \eta qx}b_{q \eta}^{\dagger}\ee^{-aq/2}
\end{equation}
and \(\varphi_{\eta}(x) = \big[\varphi_{\eta}^\dagger(x)\big]^\dagger\),
where \(a\) is a short-distance cutoff which is required to regularize certain expressions. They have commutators \([\varphi_{\eta}(x),N_{\eta'}] = [\varphi_\eta(x),\varphi_{\eta'}(x')] = 0\) and
\begin{align}
\label{eq:comm1}
[\varphi_{\eta}(x),\varphi_{\eta'}^{\dagger}(x')] &= \frac{\delta_{\eta\eta'}}{(2\pi)^2}\sum_{q>0} \frac{1}{n_{q}}\ee^{q[\ii \eta(x-x')-a]} \\
\label{eq:comm2}
&=-\frac{\delta_{\eta\eta'}}{(2\pi)^2}\log{\left[1 - \ee^{\ii \frac{2\pi}{L_{x}}[\eta(x-x')+\ii a]}\right]}\punc.
\end{align}
In particular,
\beq[eq:varphixxcomm]
[\varphi_{\eta}(x),\varphi_{\eta'}^{\dagger}(x)] =-\frac{\delta_{\eta\eta'}}{(2\pi)^2}\log{\frac{2\pi a}{L_{x}}}\punc,
\eeq
since \(a \ll L_x\). As with \(b_{q\eta}\), we have \(\varphi_\eta\lvert 0 \rangle = 0\).

From these operators, we construct the Hermitian combination
\begin{equation}
\label{eq:chiralfields}
\phi_{\eta}(x) = \phiz_\eta + \frac{x}{L_{x}}\left(N_{\eta} + \frac{1}{2}p\right) + \varphi_{\eta}(x) + \varphi_{\eta}^{\dagger}(x) \punc.
\end{equation}
The zero-wavevector mode of \(\phi_\eta(x)\) is the (\(x\)-independent) operator
\beq
\phiz_\eta = -\frac{\eta}{2\pi}\theta_\eta
\eeq
where \(\theta_\eta\) (\(=\theta_\eta^\dagger\)) is dual to \(N_\eta\), i.e., \(\ee^{+\ii\theta_\eta}\) (the ``Klein factor'' \(F_\eta^\dagger\) \cite{vonDelft1998}) increments \(N_\eta\). Equivalently,
\beq{}
[N_\eta,\theta_{\eta'}] = -\ii \delta_{\eta\eta'} \punc,
\eeq
but with the important caveat that \(\theta_\eta\) is defined only up to multiples of \(2\pi\) (since \(N_\eta\) takes integer values) and hence the zero mode \(\phiz_\eta\) is defined only up to integers. We also require
\beq[eq:phizcommutator]
[\theta_+,\theta_-] = \ii \pi\punc,
\eeq
and that the operators \(\theta_\eta\) commute with \(\varphi_\eta(x)\).

The real-space fermionic operator \(\Psi_\eta(x)\) can be expressed in terms of the bosons through the bosonization identity\footnote{For the left-moving fermions, \(\eta = -\), our \refeq{eq:bosonizationidentity} is equal, up to normalization of the fields, to Eq.~(62) of \refcite{vonDelft1998} with \(F_\eta = \ee^{-\ii \theta_\eta}\). Our \(\phi_\eta\) is equivalent to \(\tilde{\Phi}_\eta\), defined in Section D.2 of \refcite{vonDelft1998}.}
\begin{align}
\label{eq:bosonizationidentity}
\Psi_\eta(x) &= \frac{1}{\sqrt{L_{x}}}\ee^{-\ii\theta_\eta}\ee^{{2\pi}\ii \eta (N_{\eta}-\frac{1}{2}+\frac{1}{2}p)x/{L_{x}}}\ee^{2\pi\ii \eta\varphi_{\eta}^{\dagger}(x)}\ee^{2\pi\ii \eta\varphi_{\eta}(x)} \\
\label{eq:bosonizationidentity2}
&= \frac{1}{\sqrt{2\pi a}}\ee^{2\pi\ii \eta\phi_{\eta}(x)} \punc,
\end{align}
which becomes an exact operator identity in the limit \(a\rightarrow 0\) \cite{vonDelft1998}. The two expressions can be related combining the exponentials using \refeq{eq:corollary1}. In the first, the boson operators are in normal order, while in the second, \(\ee^{2\pi\ii \eta\phi_{\eta}(x)}\) is not normal ordered and so has a nontrivial limit as \(a \rightarrow 0\).

Finally, we define the fields
\begin{align}
\label{eq:phi} \phi(x) &= \phi_+(x) + \phi_-(x) \punc,\\
\label{eq:theta} \thetai(x) &= \phi_+(x) - \phi_-(x) \punc,
\end{align}
and $\Pi(x) = -\pi\thetai'(x)$. To find their commutation relations, we use \refeq{eq:comm1} and the identity
\begin{equation}
\sum_{n=-\infty}^{\infty}\ee^{\ii n \xi}\ee^{-\lambda\lvert n\rvert} = 2\pi\sum_{m=-\infty}^{\infty}\delta_{\lambda}(\xi-2\pi m)\punc ,
\end{equation}
where
\begin{equation}
\label{eq:Lorentzian}
\delta_\lambda(x) = \frac{\lambda}{\pi}\frac{1}{\lambda^2 + x^2}
\end{equation}
is a Lorentzian of width \(\lambda\). These give \([\phi(x),\phi(x')] = [\vartheta(x),\vartheta(x')] = [\Pi(x),\Pi(x')] = 0\) and
\begin{equation}
\label{eq:ccmva}
[\phi (x),\Pi(x')] = \ii  \Sha_a(x-x') \punc,
\end{equation}
where
\beq[eq:defineSha]
\Sha_a(x) = \sum_{n=-\infty}^{\infty}\delta_a(x-nL_x)\punc.
\eeq
In the limit \(a\rightarrow 0\), \(\Sha_a\) becomes a Dirac comb (periodic delta function) with period \(L_x\). In this limit, the field \(\Pi(x)\) therefore plays the role of the canonically conjugate momentum for the bosonic field \(\phi(x)\).

Integrating \refeq{eq:ccmva} and using the boundary condition \([\phi(x),\vartheta(x)] = \frac{\ii}{2\pi}\), which follows from \refeqand{eq:varphixxcomm}{eq:phizcommutator}, we find
\beq[eq:phivarthetacomm]
[\phi(x),\vartheta(x')] = \frac{\ii}{\pi}\Theta_a(x-x')\punc,
\eeq
where \(\Theta_a\), defined by \(\Theta'_a(x)=\Sha_a(x)\) and \(\Theta_a(0) = \frac{1}{2}\), is a smoothed ceiling function,
\beq
\lim_{a \rightarrow 0} \Theta_a(x) = \left\lceil\frac{x}{L_x}\right\rceil
\eeq
for \({x}/{L_x} \not\in \mathbb{Z}\). For \(0 \le x,x' < L_x\), \(\Theta_a\) can be replaced by a unit step in this limit.

We note here two important properties of the field \(\phi\): First, \refeq{eq:chiralfields} implies it is not periodic, but rather satisfies
\begin{align}
\label{eq:phiaperiodic}
\phi(L_{x}) - \phi(0) &= p + N_{+} + N_{-}\\
&= p + \int_0^{L_x}\!\dd x\, \no{\Psi_{+}^{\dagger}\Psi_{+} + \Psi_{-}^{\dagger}\Psi_{-}}
\label{eq:phiaperiodic2}
\punc ,
\end{align}
where \refeq{eq:Neta} has been used. Second, because of the definition of the zero mode \(\phiz_\eta\), both \(\phi\) and \(\thetai\) (but not \(\Pi\)) are defined only up to global integer shifts.

\subsection{Bosonization dictionary}
\label{bosonizationdictionary}

We now use the bosonization identity to derive a `dictionary' of useful bosonization formulae, as summarized in \reftab{tab:bosonizationdictionary}. (In the table, we make the replacements \(\Psi_+ \rightarrow R\) and \(\Psi_- \rightarrow L\) to match the notation used in the main text.)

In order to bosonize bilinears (and their derivatives) that contain a single fermion species, as shown in the first three rows of \reftab{tab:bosonizationdictionary}, we consider the two-point function
\begin{align}
G_{\eta}(x,y) &= \no{\Psi_\eta^{\dagger}(x)\Psi_\eta(y)} \\ 
\label{eq:divs} &= \Psi_\eta^{\dagger}(x)\Psi_\eta(y) - \langle 0 \rvert \Psi_\eta^{\dagger}(x)\Psi_\eta(y) \lvert 0 \rangle \punc ,
\end{align}
such that
\begin{equation}
\label{eq:genfun}
\no{(\partial_{x}^{m}\Psi_\eta^{\dagger})\partial_{x}^{n}\Psi_\eta} = \lim_{y\rightarrow x}\partial_{x}^{m}\partial_{y}^{n}G_{\eta}(x,y) \punc .
\end{equation}
Products of fermion operators diverge when evaluated at coinciding points, and so \refeq{eq:divs} is valid only for \(x \neq y\). The limit in \refeq{eq:genfun} is well defined, however, because normal-ordering removes the divergences; this technique is referred to as ``point splitting'' \cite{Senechal2004}.

We now express $G_\eta(x,y)$ in terms of boson operators. To do so, we start from \refeq{eq:bosonizationidentity} and use 
\refeqand{eq:corollary2}{eq:comm2} to exchange $\ee^{-2\pi\ii \eta\varphi_{\eta}(x)}$ and $\ee^{2\pi\ii \eta\varphi_{\eta}^{\dagger}(y)}$, with the result
\begin{equation}
\Psi_\eta^{\dagger}(x)\Psi_\eta(y) = \frac{1}{L_x} \frac{\bno{\ee^{-2\pi\ii \eta[\phi_{\eta}(x)-\phi_{\eta}(y)]}}}{1-\ee^{2\pi \ii \eta(x-y)/L_x}} \punc ,
\end{equation}
in the limit \(a \rightarrow 0\). The expression in the numerator is shorthand for
\begin{multline}
\label{eq:bosonno}
\bno{\ee^{-2\pi\ii \eta[\phi_{\eta}(x)-\phi_{\eta}(y)]}} \equiv \ee^{-{2\pi}\ii \eta (N_{\eta}-\frac{1}{2}+\frac{1}{2}p)(x-y)/{L_{x}}}
\\\times\ee^{-2\pi\ii \eta[\varphi_{\eta}^{\dagger}(x)-\varphi_{\eta}^{\dagger}(y)]}
\ee^{-2\pi\ii \eta[\varphi_{\eta}(x)-\varphi_{\eta}(y)]}\punc,
\end{multline}
where the operator exponentials have been rearranged into normal order, with all bosonic annihilation operators \(\varphi_\eta\) to the right of all creation operators \(\varphi_\eta^\dagger\). Because \(\varphi_\eta\lvert 0 \rangle = N_\eta \lvert 0 \rangle = 0\), the expectation value of this operator in the ground state is
\beq
\langle 0 \rvert \bno{\ee^{-2\pi\ii \eta[\phi_{\eta}(x)-\phi_{\eta}(y)]}}\lvert 0 \rangle = \ee^{\pi\ii \eta (1-p)(x-y)/{L_{x}}}
\eeq
and so we obtain
\begin{equation}
\label{eq:frfinal}
G_{\eta}(x,y) = \frac{1}{L_x}\frac{\bno{\ee^{-2\pi\ii \eta[\phi_{\eta}(x)-\phi_{\eta}(y)]}} - \ee^{\pi\ii \eta (1-p)(x-y)/{L_{x}}}}{1-\ee^{2\pi \ii \eta(x-y)/L_x}} \punc .
\end{equation}

We can now use \refeq{eq:genfun} to express local bilinears of fermions in terms of the bosonic operators. The simplest are the densities of right- and left-movers relative to the ground state, given by
\begin{align}
\no{\Psi_\eta^{\dagger}(x) \Psi_\eta(x)} &= \lim_{\epsilon \rightarrow 0}G_{\eta}(x,x+\epsilon) \\
&= \lim_{\epsilon\rightarrow 0}\frac{\bno{\ee^{2\pi\ii \eta\epsilon\partial_{x}\phi_{\eta}(x)}}-\ee^{-\pi\ii \eta (1-p)\epsilon/{L_{x}}}}{2\pi\ii \eta \epsilon}\punc,
\end{align}
where
\begin{multline}
\bno{\ee^{2\pi\ii \eta\epsilon\partial_{x}\phi_{\eta}(x)}} \equiv \ee^{{2\pi}\ii \eta\epsilon (N_{\eta}-\frac{1}{2}+\frac{1}{2}p)/{L_{x}}}
\\\times\ee^{2\pi\ii \eta\epsilon\partial_x\varphi_{\eta}^{\dagger}(x)}
\ee^{2\pi\ii \eta\epsilon\partial_x\varphi_{\eta}(x)}\punc.
\end{multline}
Expanding the exponentials and collecting terms gives
\begin{align}
\no{\Psi_\eta^{\dagger}(x) \Psi_\eta(x)} &= \frac{1}{L_x}N_\eta + \partial_x \varphi_\eta(x) + \partial_x \varphi_\eta^\dagger(x)\\
&= \partial_x \phi_\eta(x) - \frac{1}{2L_x}p
\label{eq:dxphir}
\punc .
\end{align}
Hence, using \refeqand{eq:phi}{eq:theta}, we find
\begin{align}
\label{eq:dxphi}
\no{\Psi_+^{\dagger}\Psi_+ + \Psi_-^{\dagger}\Psi_-} &= \partial_{x}\phi - \frac{1}{L_x}p \\
\label{eq:Pi}
\no{\Psi_+^{\dagger}\Psi_+ - \Psi_-^{\dagger}\Psi_-} &= -\frac{1}{\pi}\Pi \punc .
\end{align}
Note that \refeq{eq:dxphi} is consistent with \refeq{eq:phiaperiodic2}.

Similarly, for the terms with first derivatives, we compute
\begin{equation}
\label{eq:hamnoint0}
\frac{1}{2\ii}\no{\Psi_\eta^{\dagger}\partial_{x}\Psi_\eta - (\partial_{x}\Psi_\eta^{\dagger})\Psi_\eta} = \eta\pi\bno{(\partial_{x}\phi_{\eta})^{2}} - \frac{\pi \eta}{4L_x^2}p \punc ,
\end{equation}
which implies
\begin{multline}
\label{eq:hamnoint}
\frac{1}{2\ii}\no{\Psi_+^{\dagger}\partial_{x}\Psi_+ - (\partial_{x}\Psi_+^{\dagger})\Psi_+ - \Psi_-^{\dagger}\partial_{x}\Psi_- + (\partial_{x}\Psi_-^{\dagger})\Psi_-} \\
= \frac{\pi}{2}\bno{(\partial_{x}\phi)^{2}+\frac{1}{\pi^2}\Pi^{2}} - \frac{\pi}{2L_x^2}p \punc .
\end{multline}
On the right-hand side of \refeqand{eq:hamnoint0}{eq:hamnoint}, normal ordering simply amounts to adding a constant,
\beq[eq:bnophi2]
\begin{aligned}
\bno{(\partial_{x}\phi_{\eta})(\partial_{x}\phi_{\eta'})} &= (\partial_{x}\phi_{\eta})(\partial_{x}\phi_{\eta'}) - \left[\partial_x\varphi_\eta(x),\partial_x\varphi_{\eta'}^\dagger(x)\right]\\
&= (\partial_{x}\phi_{\eta})(\partial_{x}\phi_{\eta'}) - \frac{\delta_{\eta\eta'}}{(2\pi a)^2}\punc.
\end{aligned}
\eeq

We also require a bosonized expression for the combination \(\Psi_-^\dagger(x) \Psi_+(x)\), which mixes right- and left-moving fermions. This can be found using the second form of the bosonization identity, \refeq{eq:bosonizationidentity2}, which gives
\beq
\Psi_-^\dagger(x)\Psi_+(x) = \frac{1}{2\pi a} \ee^{2\pi \ii \phi_-(x)}\ee^{2\pi \ii \phi_+(x)}\punc.
\eeq
Using \refeq{eq:corollary1} and
\beq{}
[\phi_-(x),\phi_+(x)] = -\frac{1}{(2\pi)^2}[\theta_-,\theta_+] = \frac{\ii}{4\pi}
\eeq
to combine the exponentials, we find simply
\beq
\Psi_-^\dagger(x)\Psi_+(x) = \frac{1}{2\pi\ii a}\ee^{2\pi \ii \phi(x)}\punc,
\eeq
and so
\begin{align}
\label{eq:RLsin}
\Psi_+^\dagger \Psi_- + \Psi_-^\dagger \Psi_+ &= \frac{1}{\pi a}\sin (2\pi \phi)\\
\label{eq:RLcos}
\Psi_+^\dagger \Psi_- - \Psi_-^\dagger \Psi_+ &= \frac{\ii}{\pi a}\cos (2\pi \phi)\punc.
\end{align}

Note that, as expected, all of the bosonic expressions are invariant under integer shifts of \(\phi\).

\section{Path integral and regularization}
\label{regularization}

The fields \(\phi\) and \(\Pi\) constructed in \refapp{bosonization} have commutation relation [\refeq{eq:ccmva}]
\begin{equation}
[\phi (x),\Pi(x')] = \ii \Sha_a(x-x') \punc,
\end{equation}
where \(\Sha_a\) is a periodic Lorentzian of width \(a\) and period \(L_x\) and \(a\) is a small cutoff with dimensions of length, originally introduced in \refeq{eq:chiralladderoperators}. Strictly speaking, the fields therefore become conjugate position and momentum variables only in the limit \(a \rightarrow 0\), where \(\Sha_a\) becomes a periodic delta function.

Rather than taking the limit \(a\rightarrow0\) before mapping to a path-integral representation, we keep \(a\) as a short length scale. As we show below, this is sufficient to regularize the bosonic field theory, and so we use \(a\) rather than introducing a separate short-distance cutoff. (Note that \(a\) is not the lattice spacing, which we set to \(1\) in the microscopic theory.) Although including \(a\) breaks the symmetry between \(x\) and \(y\) in the path integral, it correctly gives isotropic correlations for distances much larger than \(a\) [see, e.g., \refeq{eq:phicorrelation}].

To make this explicit, we introduce a periodic field \(\Phii(x)\) through
\begin{equation}
\label{eq:Phidef}
\phi(x) = (\Sha_{a}*\Phii)(x) + \frac{x\Phi_y}{L_x}\punc ,
\end{equation}
where \(*\) denotes convolution,
\beq
(\Sha_{a}*\Phii)(x) = \int_{0}^{L_x} \dd x' \, \Sha_a(x-x')\Phii(x')
\eeq
and the second term in \refeq{eq:Phidef} gives the correct boundary condition, \refeq{eq:phibc}.
Comparison with \refeq{eq:ccmva} shows that the operator \(\chi\) has commutator
\begin{equation}
\label{eq:ccmvPhi}
[\Phii(x),\Pi(x')] = \ii \Sha(x-x') \punc,
\end{equation}
where \(\Sha\) is a (zero-width) periodic delta function, and so is exactly the conjugate field to the momentum \(\Pi\). [From \refeq{eq:phiydphi}, it follows that \([\Phi_y,\Pi(x)] = 0\).]

The mapping from the Hamiltonian to a path integral, described in \refsec{action}, should then be performed using complete sets of eigenstates of \(\Phii\), leading to an action
\begin{equation}
S[\Phii] = \frac{\anis\pi}{2}\int_0^{L_x} \!\dd x \int_0^{L_y} \!\dd y \, \left[(\partial_{x}\phi)^{2} + \frac{1}{\anis^2}(\partial_{y}\Phii)^{2}\right] \punc,
\end{equation}
where \(\phi\) is given in terms of \(\Phii\) by \refeq{eq:Phidef}. This should be understood as the regularized version of the action given in the main text, \refeq{eq:sphi}.

We use this action to calculate regularized correlation functions for separations much smaller than the system size. To do so, we define the Fourier transform \(\tilde{\chi}(\bm{k})\) of $\Phii$ by
\begin{equation}
\label{eq:ftPhi}
\Phii(\rv) = \int \frac{d^{2}\bm{k}}{2\pi}\ee^{-\ii\bm{k}\cdot\rv}\tilde{\Phii}(\bm{k})
\punc,
\end{equation}
where we have taken the thermodynamic limit (\(L_x,L_y\rightarrow\infty\)) and neglected boundary effects. The regularized action can then be written as
\begin{equation}
S[\tilde{\Phii}] = \frac{\anis\pi}{2}\int \dd^{2}\bm{k}\left[k_{x}^{2}\tilde{\delta}_{a}(k_x)^{2} + \frac{k_{y}^{2}}{\anis^2}\right]\left\lvert \tilde{\Phii}(\bm{k})\right\rvert^{2} \punc ,
\end{equation}
where $\tilde{\delta}_{a}(k) = \int_{-\infty}^{\infty}\dd x \, \ee^{\ii kx}\delta_{a}(x)= \ee^{-\lvert k\rvert a}$, and so the correlation functions of $\tilde{\Phii}$ are
\begin{equation}
\langle \tilde{\Phii}(\bm{k})\tilde{\Phii}(-\bm{k'})\rangle = \frac{1}{\anis\pi}\frac{\delta^{2}(\bm{k}-\bm{k}')}{k_{x}^{2}\tilde{\delta}_{a}(k_{x})^{2} + {k_{y}^{2}}/{\anis^2}} \punc .
\end{equation}
Although the correlation function \(\langle [\Phii(\rv) - \Phii(\bm{r}') ]^2\rangle\) diverges in the UV (in fact linearly, rather than logarithmically as when \(a=0\)), the correlation functions of $\phi(\rv)$ are finite. From \refeqand{eq:Phidef}{eq:ftPhi}, we have (for \(L_x \rightarrow\infty\))
\begin{equation}
\phi(\rv) = \int \frac{d^{2}\bm{k}}{2\pi}\ee^{-\ii\bm{k}\cdot\rv}\tilde{\delta}_{a}(k_{x})\tilde{\Phii}(\bm{k}) \punc ,
\end{equation}
and hence [\refeq{eq:asymptoticcorrelations}]
\begin{equation}
\label{eq:phicorrelation}
\left\langle \left[\phi(\rv) - \phi(\bm{r}') \right]^2\right\rangle =\frac{1}{\pi^2} \log\left( \frac{\lvert \tilde{\bm{r}} - \tilde{\bm{r}}' \rvert}{a}\right) + O\left(\frac{\lvert\tilde{\rv}-\tilde{\rv}'\rvert}{a}\right)^{-1} \punc,
\end{equation}
where $\tilde{\bm{r}}=(x,\anis y)$. The cutoff $a$ therefore regularizes $\phi$ correlators.
The UV cutoff is typically inserted by hand when calculating $\phi$ correlators \cite{Chalker2017}, leading to the same asymptotic form as \refeq{eq:phicorrelation} but with leading-order corrections $O(a^{0})$ rather than \(O(a^1)\).

We note also that in the short-distance limit, \(\lvert \rv-\rv' \rvert \ll a\),
\beq
\left\langle \left[\phi(\rv) - \phi(\bm{r}') \right]^2\right\rangle = O\left(\frac{\lvert \rv -\rv' \rvert}{a}\right)\punc,
\eeq
meaning that the regularized path integral is dominated by H\"older-continuous field configurations \(\phi(\rv)\).

\bibliography{dimersbibliography}

\begin{thebibliography}{58}%
\makeatletter
\providecommand \@ifxundefined [1]{%
 \@ifx{#1\undefined}
}%
\providecommand \@ifnum [1]{%
 \ifnum #1\expandafter \@firstoftwo
 \else \expandafter \@secondoftwo
 \fi
}%
\providecommand \@ifx [1]{%
 \ifx #1\expandafter \@firstoftwo
 \else \expandafter \@secondoftwo
 \fi
}%
\providecommand \natexlab [1]{#1}%
\providecommand \enquote  [1]{``#1''}%
\providecommand \bibnamefont  [1]{#1}%
\providecommand \bibfnamefont [1]{#1}%
\providecommand \citenamefont [1]{#1}%
\providecommand \href@noop [0]{\@secondoftwo}%
\providecommand \href [0]{\begingroup \@sanitize@url \@href}%
\providecommand \@href[1]{\@@startlink{#1}\@@href}%
\providecommand \@@href[1]{\endgroup#1\@@endlink}%
\providecommand \@sanitize@url [0]{\catcode `\\12\catcode `\$12\catcode
  `\&12\catcode `\#12\catcode `\^12\catcode `\_12\catcode `\%12\relax}%
\providecommand \@@startlink[1]{}%
\providecommand \@@endlink[0]{}%
\providecommand \url  [0]{\begingroup\@sanitize@url \@url }%
\providecommand \@url [1]{\endgroup\@href {#1}{\urlprefix }}%
\providecommand \urlprefix  [0]{URL }%
\providecommand \Eprint [0]{\href }%
\providecommand \doibase [0]{https://doi.org/}%
\providecommand \selectlanguage [0]{\@gobble}%
\providecommand \bibinfo  [0]{\@secondoftwo}%
\providecommand \bibfield  [0]{\@secondoftwo}%
\providecommand \translation [1]{[#1]}%
\providecommand \BibitemOpen [0]{}%
\providecommand \bibitemStop [0]{}%
\providecommand \bibitemNoStop [0]{.\EOS\space}%
\providecommand \EOS [0]{\spacefactor3000\relax}%
\providecommand \BibitemShut  [1]{\csname bibitem#1\endcsname}%
\let\auto@bib@innerbib\@empty
\bibitem [{\citenamefont {Henley}(2010)}]{Henley2010}%
  \BibitemOpen
  \bibfield  {author} {\bibinfo {author} {\bibfnamefont {C.~L.}\ \bibnamefont
  {Henley}},\ }\bibfield  {title} {\bibinfo {title} {The
  {\textquotedblleft}{C}oulomb phase{\textquotedblright} in frustrated
  systems},\ }\href {https://doi.org/10.1146/annurev-conmatphys-070909-104138}
  {\bibfield  {journal} {\bibinfo  {journal} {Annual Review of Condensed Matter
  Physics}\ }\textbf {\bibinfo {volume} {1}},\ \bibinfo {pages} {179} (\bibinfo
  {year} {2010})}\BibitemShut {NoStop}%
\bibitem [{\citenamefont {Anderson}(1956)}]{Anderson1956}%
  \BibitemOpen
  \bibfield  {author} {\bibinfo {author} {\bibfnamefont {P.~W.}\ \bibnamefont
  {Anderson}},\ }\bibfield  {title} {\bibinfo {title} {Ordering and
  antiferromagnetism in ferrites},\ }\href
  {https://doi.org/10.1103/PhysRev.102.1008} {\bibfield  {journal} {\bibinfo
  {journal} {Phys. Rev.}\ }\textbf {\bibinfo {volume} {102}},\ \bibinfo {pages}
  {1008} (\bibinfo {year} {1956})}\BibitemShut {NoStop}%
\bibitem [{\citenamefont {Youngblood}\ and\ \citenamefont
  {Axe}(1981)}]{Youngblood1981}%
  \BibitemOpen
  \bibfield  {author} {\bibinfo {author} {\bibfnamefont {R.~W.}\ \bibnamefont
  {Youngblood}}\ and\ \bibinfo {author} {\bibfnamefont {J.~D.}\ \bibnamefont
  {Axe}},\ }\bibfield  {title} {\bibinfo {title} {Polarization fluctuations in
  ferroelectric models},\ }\href {https://doi.org/10.1103/PhysRevB.23.232}
  {\bibfield  {journal} {\bibinfo  {journal} {Phys. Rev. B}\ }\textbf {\bibinfo
  {volume} {23}},\ \bibinfo {pages} {232} (\bibinfo {year} {1981})}\BibitemShut
  {NoStop}%
\bibitem [{\citenamefont {Bl{\"o}te}\ and\ \citenamefont
  {Hilhorst}(1982)}]{Blote1982}%
  \BibitemOpen
  \bibfield  {author} {\bibinfo {author} {\bibfnamefont {H.~W.~J.}\
  \bibnamefont {Bl{\"o}te}}\ and\ \bibinfo {author} {\bibfnamefont {H.~J.}\
  \bibnamefont {Hilhorst}},\ }\bibfield  {title} {\bibinfo {title} {Roughening
  transitions and the zero-temperature triangular {I}sing antiferromagnet},\
  }\href {https://doi.org/10.1088/0305-4470/15/11/011} {\bibfield  {journal}
  {\bibinfo  {journal} {Journal of Physics A: Mathematical and General}\
  }\textbf {\bibinfo {volume} {15}},\ \bibinfo {pages} {L631} (\bibinfo {year}
  {1982})}\BibitemShut {NoStop}%
\bibitem [{\citenamefont {Zeng}\ and\ \citenamefont {Henley}(1997)}]{Zeng1997}%
  \BibitemOpen
  \bibfield  {author} {\bibinfo {author} {\bibfnamefont {C.}~\bibnamefont
  {Zeng}}\ and\ \bibinfo {author} {\bibfnamefont {C.~L.}\ \bibnamefont
  {Henley}},\ }\bibfield  {title} {\bibinfo {title} {Zero-temperature phase
  transitions of an antiferromagnetic {I}sing model of general spin on a
  triangular lattice},\ }\href {https://doi.org/10.1103/PhysRevB.55.14935}
  {\bibfield  {journal} {\bibinfo  {journal} {Phys. Rev. B}\ }\textbf {\bibinfo
  {volume} {55}},\ \bibinfo {pages} {14935} (\bibinfo {year}
  {1997})}\BibitemShut {NoStop}%
\bibitem [{\citenamefont {Kondev}\ and\ \citenamefont
  {Henley}(1996)}]{Kondev1996}%
  \BibitemOpen
  \bibfield  {author} {\bibinfo {author} {\bibfnamefont {J.}~\bibnamefont
  {Kondev}}\ and\ \bibinfo {author} {\bibfnamefont {C.~L.}\ \bibnamefont
  {Henley}},\ }\bibfield  {title} {\bibinfo {title} {{K}ac--{M}oody symmetries
  of critical ground states},\ }\href
  {https://doi.org/https://doi.org/10.1016/0550-3213(96)00064-8} {\bibfield
  {journal} {\bibinfo  {journal} {Nuclear Physics B}\ }\textbf {\bibinfo
  {volume} {464}},\ \bibinfo {pages} {540 } (\bibinfo {year}
  {1996})}\BibitemShut {NoStop}%
\bibitem [{\citenamefont {Castelnovo}\ \emph {et~al.}(2012)\citenamefont
  {Castelnovo}, \citenamefont {Moessner},\ and\ \citenamefont
  {Sondhi}}]{Castelnovo2012}%
  \BibitemOpen
  \bibfield  {author} {\bibinfo {author} {\bibfnamefont {C.}~\bibnamefont
  {Castelnovo}}, \bibinfo {author} {\bibfnamefont {R.}~\bibnamefont
  {Moessner}},\ and\ \bibinfo {author} {\bibfnamefont {S.~L.}\ \bibnamefont
  {Sondhi}},\ }\bibfield  {title} {\bibinfo {title} {Spin ice,
  fractionalization, and topological order},\ }\href
  {https://doi.org/10.1146/annurev-conmatphys-020911-125058} {\bibfield
  {journal} {\bibinfo  {journal} {Annual Review of Condensed Matter Physics}\
  }\textbf {\bibinfo {volume} {3}},\ \bibinfo {pages} {35} (\bibinfo {year}
  {2012})}\BibitemShut {NoStop}%
\bibitem [{\citenamefont {Chalker}(2017)}]{Chalker2017}%
  \BibitemOpen
  \bibfield  {author} {\bibinfo {author} {\bibfnamefont {J.~T.}\ \bibnamefont
  {Chalker}},\ }\bibfield  {title} {\bibinfo {title} {Spin liquids and
  frustrated magnetism},\ }in\ \href@noop {} {\emph {\bibinfo {booktitle}
  {Topological Aspects of Condensed Matter Physics}}},\ Vol.\ \bibinfo {volume}
  {103},\ \bibinfo {editor} {edited by\ \bibinfo {editor} {\bibfnamefont
  {C.}~\bibnamefont {Chamon}}, \bibinfo {editor} {\bibfnamefont
  {M.}~\bibnamefont {Goerbig}}, \bibinfo {editor} {\bibfnamefont
  {R.}~\bibnamefont {Moessner}},\ and\ \bibinfo {editor} {\bibfnamefont
  {L.}~\bibnamefont {Cugliandolo}}}\ (\bibinfo  {publisher} {Oxford University
  Press},\ \bibinfo {year} {2017})\ \bibinfo {note} {{L}ecture notes of the
  {L}es {H}ouches {S}ummer {S}chool, {A}ugust 2014}\BibitemShut {NoStop}%
\bibitem [{\citenamefont {Huse}\ \emph {et~al.}(2003)\citenamefont {Huse},
  \citenamefont {Krauth}, \citenamefont {Moessner},\ and\ \citenamefont
  {Sondhi}}]{Huse2003}%
  \BibitemOpen
  \bibfield  {author} {\bibinfo {author} {\bibfnamefont {D.~A.}\ \bibnamefont
  {Huse}}, \bibinfo {author} {\bibfnamefont {W.}~\bibnamefont {Krauth}},
  \bibinfo {author} {\bibfnamefont {R.}~\bibnamefont {Moessner}},\ and\
  \bibinfo {author} {\bibfnamefont {S.~L.}\ \bibnamefont {Sondhi}},\ }\bibfield
   {title} {\bibinfo {title} {{C}oulomb and liquid dimer models in three
  dimensions},\ }\href {https://doi.org/10.1103/PhysRevLett.91.167004}
  {\bibfield  {journal} {\bibinfo  {journal} {Phys. Rev. Lett.}\ }\textbf
  {\bibinfo {volume} {91}},\ \bibinfo {pages} {167004} (\bibinfo {year}
  {2003})}\BibitemShut {NoStop}%
\bibitem [{\citenamefont {Alet}\ \emph {et~al.}(2005)\citenamefont {Alet},
  \citenamefont {Jacobsen}, \citenamefont {Misguich}, \citenamefont {Pasquier},
  \citenamefont {Mila},\ and\ \citenamefont {Troyer}}]{Alet2005}%
  \BibitemOpen
  \bibfield  {author} {\bibinfo {author} {\bibfnamefont {F.}~\bibnamefont
  {Alet}}, \bibinfo {author} {\bibfnamefont {J.~L.}\ \bibnamefont {Jacobsen}},
  \bibinfo {author} {\bibfnamefont {G.}~\bibnamefont {Misguich}}, \bibinfo
  {author} {\bibfnamefont {V.}~\bibnamefont {Pasquier}}, \bibinfo {author}
  {\bibfnamefont {F.}~\bibnamefont {Mila}},\ and\ \bibinfo {author}
  {\bibfnamefont {M.}~\bibnamefont {Troyer}},\ }\bibfield  {title} {\bibinfo
  {title} {Interacting classical dimers on the square lattice},\ }\href
  {https://doi.org/10.1103/PhysRevLett.94.235702} {\bibfield  {journal}
  {\bibinfo  {journal} {Phys. Rev. Lett.}\ }\textbf {\bibinfo {volume} {94}},\
  \bibinfo {pages} {235702} (\bibinfo {year} {2005})}\BibitemShut {NoStop}%
\bibitem [{\citenamefont {Alet}\ \emph
  {et~al.}(2006{\natexlab{a}})\citenamefont {Alet}, \citenamefont {Misguich},
  \citenamefont {Pasquier}, \citenamefont {Moessner},\ and\ \citenamefont
  {Jacobsen}}]{Alet2006}%
  \BibitemOpen
  \bibfield  {author} {\bibinfo {author} {\bibfnamefont {F.}~\bibnamefont
  {Alet}}, \bibinfo {author} {\bibfnamefont {G.}~\bibnamefont {Misguich}},
  \bibinfo {author} {\bibfnamefont {V.}~\bibnamefont {Pasquier}}, \bibinfo
  {author} {\bibfnamefont {R.}~\bibnamefont {Moessner}},\ and\ \bibinfo
  {author} {\bibfnamefont {J.~L.}\ \bibnamefont {Jacobsen}},\ }\bibfield
  {title} {\bibinfo {title} {Unconventional continuous phase transition in a
  three-dimensional dimer model},\ }\href
  {https://doi.org/10.1103/PhysRevLett.97.030403} {\bibfield  {journal}
  {\bibinfo  {journal} {Phys. Rev. Lett.}\ }\textbf {\bibinfo {volume} {97}},\
  \bibinfo {pages} {030403} (\bibinfo {year} {2006}{\natexlab{a}})}\BibitemShut
  {NoStop}%
\bibitem [{\citenamefont {{Henley}}(1997)}]{Henley1997}%
  \BibitemOpen
  \bibfield  {author} {\bibinfo {author} {\bibfnamefont {C.~L.}\ \bibnamefont
  {{Henley}}},\ }\bibfield  {title} {\bibinfo {title} {{Relaxation time for a
  dimer covering with height representation}},\ }\href@noop {} {\bibfield
  {journal} {\bibinfo  {journal} {Journal of Statistical Physics}\ }\textbf
  {\bibinfo {volume} {89}},\ \bibinfo {pages} {483} (\bibinfo {year}
  {1997})}\BibitemShut {NoStop}%
\bibitem [{\citenamefont {Alet}\ \emph
  {et~al.}(2006{\natexlab{b}})\citenamefont {Alet}, \citenamefont {Ikhlef},
  \citenamefont {Jacobsen}, \citenamefont {Misguich},\ and\ \citenamefont
  {Pasquier}}]{Alet2006b}%
  \BibitemOpen
  \bibfield  {author} {\bibinfo {author} {\bibfnamefont {F.}~\bibnamefont
  {Alet}}, \bibinfo {author} {\bibfnamefont {Y.}~\bibnamefont {Ikhlef}},
  \bibinfo {author} {\bibfnamefont {J.~L.}\ \bibnamefont {Jacobsen}}, \bibinfo
  {author} {\bibfnamefont {G.}~\bibnamefont {Misguich}},\ and\ \bibinfo
  {author} {\bibfnamefont {V.}~\bibnamefont {Pasquier}},\ }\bibfield  {title}
  {\bibinfo {title} {Classical dimers with aligning interactions on the square
  lattice},\ }\href {https://doi.org/10.1103/PhysRevE.74.041124} {\bibfield
  {journal} {\bibinfo  {journal} {Phys. Rev. E}\ }\textbf {\bibinfo {volume}
  {74}},\ \bibinfo {pages} {041124} (\bibinfo {year}
  {2006}{\natexlab{b}})}\BibitemShut {NoStop}%
\bibitem [{\citenamefont {Kenyon}(2001)}]{Kenyon2001}%
  \BibitemOpen
  \bibfield  {author} {\bibinfo {author} {\bibfnamefont {R.}~\bibnamefont
  {Kenyon}},\ }\bibfield  {title} {\bibinfo {title} {Dominos and the {G}aussian
  free field},\ }\href {https://doi.org/10.1214/aop/1015345599} {\bibfield
  {journal} {\bibinfo  {journal} {Ann. Probab.}\ }\textbf {\bibinfo {volume}
  {29}},\ \bibinfo {pages} {1128} (\bibinfo {year} {2001})}\BibitemShut
  {NoStop}%
\bibitem [{\citenamefont {{de Tilière}}(2007)}]{deTiliere2007}%
  \BibitemOpen
  \bibfield  {author} {\bibinfo {author} {\bibfnamefont {B.}~\bibnamefont {{de
  Tilière}}},\ }\bibfield  {title} {\bibinfo {title} {Scaling limit of
  isoradial dimer models and the case of triangular quadri-tilings},\ }\href
  {https://doi.org/https://doi.org/10.1016/j.anihpb.2006.10.002} {\bibfield
  {journal} {\bibinfo  {journal} {Annales de l'Institut Henri Poincare (B)
  Probability and Statistics}\ }\textbf {\bibinfo {volume} {43}},\ \bibinfo
  {pages} {729} (\bibinfo {year} {2007})}\BibitemShut {NoStop}%
\bibitem [{\citenamefont {Giuliani}\ \emph {et~al.}(2015)\citenamefont
  {Giuliani}, \citenamefont {Mastropietro},\ and\ \citenamefont
  {Toninelli}}]{Giuliani2015}%
  \BibitemOpen
  \bibfield  {author} {\bibinfo {author} {\bibfnamefont {A.}~\bibnamefont
  {Giuliani}}, \bibinfo {author} {\bibfnamefont {V.}~\bibnamefont
  {Mastropietro}},\ and\ \bibinfo {author} {\bibfnamefont {F.}~\bibnamefont
  {Toninelli}},\ }\bibfield  {title} {\bibinfo {title} {Height fluctuations in
  non-integrable classical dimers},\ }\href
  {https://doi.org/10.1209/0295-5075/109/60004} {\bibfield  {journal} {\bibinfo
   {journal} {{EPL} (Europhysics Letters)}\ }\textbf {\bibinfo {volume}
  {109}},\ \bibinfo {pages} {60004} (\bibinfo {year} {2015})}\BibitemShut
  {NoStop}%
\bibitem [{\citenamefont {Giuliani}\ \emph
  {et~al.}(2017{\natexlab{a}})\citenamefont {Giuliani}, \citenamefont
  {Mastropietro},\ and\ \citenamefont {Toninelli}}]{Giuliani2017}%
  \BibitemOpen
  \bibfield  {author} {\bibinfo {author} {\bibfnamefont {A.}~\bibnamefont
  {Giuliani}}, \bibinfo {author} {\bibfnamefont {V.}~\bibnamefont
  {Mastropietro}},\ and\ \bibinfo {author} {\bibfnamefont {F.~L.}\ \bibnamefont
  {Toninelli}},\ }\bibfield  {title} {\bibinfo {title} {{Height fluctuations in
  interacting dimers}},\ }\href {https://doi.org/10.1214/15-AIHP710} {\bibfield
   {journal} {\bibinfo  {journal} {Annales de l'Institut Henri Poincaré,
  Probabilités et Statistiques}\ }\textbf {\bibinfo {volume} {53}},\ \bibinfo
  {pages} {98 } (\bibinfo {year} {2017}{\natexlab{a}})}\BibitemShut {NoStop}%
\bibitem [{\citenamefont {Cardy}(1996)}]{Cardy1996}%
  \BibitemOpen
  \bibfield  {author} {\bibinfo {author} {\bibfnamefont {J.}~\bibnamefont
  {Cardy}},\ }\href
  {https://www.amazon.com/Scaling-Renormalization-Statistical-Physics-Cambridge-ebook/dp/B01LYX582C?SubscriptionId=0JYN1NVW651KCA56C102&tag=techkie-20&linkCode=xm2&camp=2025&creative=165953&creativeASIN=B01LYX582C}
  {\emph {\bibinfo {title} {Scaling and Renormalization in Statistical Physics
  (Cambridge Lecture Notes in Physics)}}}\ (\bibinfo  {publisher} {Cambridge
  University Press},\ \bibinfo {year} {1996})\BibitemShut {NoStop}%
\bibitem [{\citenamefont {Berezinskii}(1971)}]{Berezinskii1971}%
  \BibitemOpen
  \bibfield  {author} {\bibinfo {author} {\bibfnamefont {V.~L.}\ \bibnamefont
  {Berezinskii}},\ }\bibfield  {title} {\bibinfo {title} {Destruction of
  long-range order in one-dimensional and two-dimensional systems having a
  continuous symmetry group {I}. {C}lassical systems},\ }\href@noop {}
  {\bibfield  {journal} {\bibinfo  {journal} {Soviet Physics JETP}\ }\textbf
  {\bibinfo {volume} {32}},\ \bibinfo {pages} {493} (\bibinfo {year}
  {1971})}\BibitemShut {NoStop}%
\bibitem [{\citenamefont {Kosterlitz}\ and\ \citenamefont
  {Thouless}(1973)}]{Kosterlitz1973}%
  \BibitemOpen
  \bibfield  {author} {\bibinfo {author} {\bibfnamefont {J.~M.}\ \bibnamefont
  {Kosterlitz}}\ and\ \bibinfo {author} {\bibfnamefont {D.~J.}\ \bibnamefont
  {Thouless}},\ }\bibfield  {title} {\bibinfo {title} {Ordering, metastability
  and phase transitions in two-dimensional systems},\ }\href
  {https://doi.org/10.1088/0022-3719/6/7/010} {\bibfield  {journal} {\bibinfo
  {journal} {Journal of Physics C: Solid State Physics}\ }\textbf {\bibinfo
  {volume} {6}},\ \bibinfo {pages} {1181} (\bibinfo {year} {1973})}\BibitemShut
  {NoStop}%
\bibitem [{\citenamefont {Kosterlitz}(2016)}]{Kosterlitz2016}%
  \BibitemOpen
  \bibfield  {author} {\bibinfo {author} {\bibfnamefont {J.~M.}\ \bibnamefont
  {Kosterlitz}},\ }\bibfield  {title} {\bibinfo {title} {Kosterlitz--{T}houless
  physics: a review of key issues},\ }\href
  {https://doi.org/10.1088/0034-4885/79/2/026001} {\bibfield  {journal}
  {\bibinfo  {journal} {Reports on Progress in Physics}\ }\textbf {\bibinfo
  {volume} {79}},\ \bibinfo {pages} {026001} (\bibinfo {year}
  {2016})}\BibitemShut {NoStop}%
\bibitem [{\citenamefont {Wilkins}\ and\ \citenamefont
  {Powell}(2020)}]{Wilkins2020}%
  \BibitemOpen
  \bibfield  {author} {\bibinfo {author} {\bibfnamefont {N.}~\bibnamefont
  {Wilkins}}\ and\ \bibinfo {author} {\bibfnamefont {S.}~\bibnamefont
  {Powell}},\ }\bibfield  {title} {\bibinfo {title} {Interacting double dimer
  model on the square lattice},\ }\href
  {https://doi.org/10.1103/PhysRevB.102.174431} {\bibfield  {journal} {\bibinfo
   {journal} {Phys. Rev. B}\ }\textbf {\bibinfo {volume} {102}},\ \bibinfo
  {pages} {174431} (\bibinfo {year} {2020})}\BibitemShut {NoStop}%
\bibitem [{\citenamefont {Desai}\ \emph {et~al.}(2021)\citenamefont {Desai},
  \citenamefont {Pujari},\ and\ \citenamefont {Damle}}]{Desai2021}%
  \BibitemOpen
  \bibfield  {author} {\bibinfo {author} {\bibfnamefont {N.}~\bibnamefont
  {Desai}}, \bibinfo {author} {\bibfnamefont {S.}~\bibnamefont {Pujari}},\ and\
  \bibinfo {author} {\bibfnamefont {K.}~\bibnamefont {Damle}},\ }\bibfield
  {title} {\bibinfo {title} {Bilayer {C}oulomb phase of two-dimensional dimer
  models: Absence of power-law columnar order},\ }\href
  {https://doi.org/10.1103/PhysRevE.103.042136} {\bibfield  {journal} {\bibinfo
   {journal} {Phys. Rev. E}\ }\textbf {\bibinfo {volume} {103}},\ \bibinfo
  {pages} {042136} (\bibinfo {year} {2021})}\BibitemShut {NoStop}%
\bibitem [{\citenamefont {Kasteleyn}(1961)}]{Kasteleyn1961}%
  \BibitemOpen
  \bibfield  {author} {\bibinfo {author} {\bibfnamefont {P.~W.}\ \bibnamefont
  {Kasteleyn}},\ }\bibfield  {title} {\bibinfo {title} {The statistics of
  dimers on a lattice},\ }\href {https://doi.org/10.1016/0031-8914(61)90063-5}
  {\bibfield  {journal} {\bibinfo  {journal} {Physica}\ }\textbf {\bibinfo
  {volume} {27}},\ \bibinfo {pages} {1209} (\bibinfo {year}
  {1961})}\BibitemShut {NoStop}%
\bibitem [{\citenamefont {Temperley}\ and\ \citenamefont
  {Fisher}(1961)}]{Temperley1961}%
  \BibitemOpen
  \bibfield  {author} {\bibinfo {author} {\bibfnamefont {H.~N.~V.}\
  \bibnamefont {Temperley}}\ and\ \bibinfo {author} {\bibfnamefont {M.~E.}\
  \bibnamefont {Fisher}},\ }\bibfield  {title} {\bibinfo {title} {Dimer problem
  in statistical mechanics--an exact result},\ }\href
  {https://doi.org/10.1080/14786436108243366} {\bibfield  {journal} {\bibinfo
  {journal} {Philosophical Magazine}\ }\textbf {\bibinfo {volume} {6}},\
  \bibinfo {pages} {1061} (\bibinfo {year} {1961})}\BibitemShut {NoStop}%
\bibitem [{\citenamefont {Fisher}(1961)}]{Fisher1961}%
  \BibitemOpen
  \bibfield  {author} {\bibinfo {author} {\bibfnamefont {M.~E.}\ \bibnamefont
  {Fisher}},\ }\bibfield  {title} {\bibinfo {title} {Statistical mechanics of
  dimers on a plane lattice},\ }\href
  {https://doi.org/10.1103/PhysRev.124.1664} {\bibfield  {journal} {\bibinfo
  {journal} {Phys. Rev.}\ }\textbf {\bibinfo {volume} {124}},\ \bibinfo {pages}
  {1664} (\bibinfo {year} {1961})}\BibitemShut {NoStop}%
\bibitem [{\citenamefont {Lieb}(1967)}]{Lieb1967}%
  \BibitemOpen
  \bibfield  {author} {\bibinfo {author} {\bibfnamefont {E.~H.}\ \bibnamefont
  {Lieb}},\ }\bibfield  {title} {\bibinfo {title} {Solution of the dimer
  problem by the transfer matrix method},\ }\href
  {https://doi.org/10.1063/1.1705163} {\bibfield  {journal} {\bibinfo
  {journal} {Journal of Mathematical Physics}\ }\textbf {\bibinfo {volume}
  {8}},\ \bibinfo {pages} {2339} (\bibinfo {year} {1967})}\BibitemShut
  {NoStop}%
\bibitem [{\citenamefont {Fisher}\ and\ \citenamefont
  {Stephenson}(1963)}]{Fisher1963}%
  \BibitemOpen
  \bibfield  {author} {\bibinfo {author} {\bibfnamefont {M.~E.}\ \bibnamefont
  {Fisher}}\ and\ \bibinfo {author} {\bibfnamefont {J.}~\bibnamefont
  {Stephenson}},\ }\bibfield  {title} {\bibinfo {title} {Statistical mechanics
  of dimers on a plane lattice. {II}. {D}imer correlations and monomers},\
  }\href {https://doi.org/10.1103/PhysRev.132.1411} {\bibfield  {journal}
  {\bibinfo  {journal} {Phys. Rev.}\ }\textbf {\bibinfo {volume} {132}},\
  \bibinfo {pages} {1411} (\bibinfo {year} {1963})}\BibitemShut {NoStop}%
\bibitem [{\citenamefont {Fradkin}\ \emph {et~al.}(2004)\citenamefont
  {Fradkin}, \citenamefont {Huse}, \citenamefont {Moessner}, \citenamefont
  {Oganesyan},\ and\ \citenamefont {Sondhi}}]{Fradkin2004}%
  \BibitemOpen
  \bibfield  {author} {\bibinfo {author} {\bibfnamefont {E.}~\bibnamefont
  {Fradkin}}, \bibinfo {author} {\bibfnamefont {D.~A.}\ \bibnamefont {Huse}},
  \bibinfo {author} {\bibfnamefont {R.}~\bibnamefont {Moessner}}, \bibinfo
  {author} {\bibfnamefont {V.}~\bibnamefont {Oganesyan}},\ and\ \bibinfo
  {author} {\bibfnamefont {S.~L.}\ \bibnamefont {Sondhi}},\ }\bibfield  {title}
  {\bibinfo {title} {Bipartite {R}okhsar--{K}ivelson points and {C}antor
  deconfinement},\ }\href {https://doi.org/10.1103/PhysRevB.69.224415}
  {\bibfield  {journal} {\bibinfo  {journal} {Phys. Rev. B}\ }\textbf {\bibinfo
  {volume} {69}},\ \bibinfo {pages} {224415} (\bibinfo {year}
  {2004})}\BibitemShut {NoStop}%
\bibitem [{\citenamefont {Papanikolaou}\ \emph {et~al.}(2007)\citenamefont
  {Papanikolaou}, \citenamefont {Luijten},\ and\ \citenamefont
  {Fradkin}}]{Papanikolaou2007}%
  \BibitemOpen
  \bibfield  {author} {\bibinfo {author} {\bibfnamefont {S.}~\bibnamefont
  {Papanikolaou}}, \bibinfo {author} {\bibfnamefont {E.}~\bibnamefont
  {Luijten}},\ and\ \bibinfo {author} {\bibfnamefont {E.}~\bibnamefont
  {Fradkin}},\ }\bibfield  {title} {\bibinfo {title} {Quantum criticality,
  lines of fixed points, and phase separation in doped two-dimensional quantum
  dimer models},\ }\href {https://doi.org/10.1103/PhysRevB.76.134514}
  {\bibfield  {journal} {\bibinfo  {journal} {Phys. Rev. B}\ }\textbf {\bibinfo
  {volume} {76}},\ \bibinfo {pages} {134514} (\bibinfo {year}
  {2007})}\BibitemShut {NoStop}%
\bibitem [{\citenamefont {Tang}\ \emph {et~al.}(2011)\citenamefont {Tang},
  \citenamefont {Sandvik},\ and\ \citenamefont {Henley}}]{Tang2011}%
  \BibitemOpen
  \bibfield  {author} {\bibinfo {author} {\bibfnamefont {Y.}~\bibnamefont
  {Tang}}, \bibinfo {author} {\bibfnamefont {A.~W.}\ \bibnamefont {Sandvik}},\
  and\ \bibinfo {author} {\bibfnamefont {C.~L.}\ \bibnamefont {Henley}},\
  }\bibfield  {title} {\bibinfo {title} {Properties of resonating-valence-bond
  spin liquids and critical dimer models},\ }\href
  {https://doi.org/10.1103/PhysRevB.84.174427} {\bibfield  {journal} {\bibinfo
  {journal} {Phys. Rev. B}\ }\textbf {\bibinfo {volume} {84}},\ \bibinfo
  {pages} {174427} (\bibinfo {year} {2011})}\BibitemShut {NoStop}%
\bibitem [{\citenamefont {Fradkin}(2013)}]{Fradkin2013}%
  \BibitemOpen
  \bibfield  {author} {\bibinfo {author} {\bibfnamefont {E.}~\bibnamefont
  {Fradkin}},\ }\href@noop {} {\emph {\bibinfo {title} {Field theories of
  condensed matter physics}}}\ (\bibinfo  {publisher} {Cambridge University
  Press},\ \bibinfo {year} {2013})\BibitemShut {NoStop}%
\bibitem [{\citenamefont {Grande}\ \emph {et~al.}(2011)\citenamefont {Grande},
  \citenamefont {Salinas},\ and\ \citenamefont {da~Costa}}]{Grande2011}%
  \BibitemOpen
  \bibfield  {author} {\bibinfo {author} {\bibfnamefont {H.~L.~C.}\
  \bibnamefont {Grande}}, \bibinfo {author} {\bibfnamefont {S.~R.}\
  \bibnamefont {Salinas}},\ and\ \bibinfo {author} {\bibfnamefont {F.~A.}\
  \bibnamefont {da~Costa}},\ }\bibfield  {title} {\bibinfo {title} {Fermionic
  representation of two-dimensional dimer models},\ }\href@noop {} {\bibfield
  {journal} {\bibinfo  {journal} {Brazilian Journal of Physics}\ }\textbf
  {\bibinfo {volume} {41}},\ \bibinfo {pages} {86} (\bibinfo {year}
  {2011})}\BibitemShut {NoStop}%
\bibitem [{\citenamefont {Wilkins}\ and\ \citenamefont
  {Powell}(2021)}]{Wilkins2021}%
  \BibitemOpen
  \bibfield  {author} {\bibinfo {author} {\bibfnamefont {N.}~\bibnamefont
  {Wilkins}}\ and\ \bibinfo {author} {\bibfnamefont {S.}~\bibnamefont
  {Powell}},\ }\bibfield  {title} {\bibinfo {title} {Topological sectors, dimer
  correlations, and monomers from the transfer-matrix solution of the dimer
  model},\ }\href {https://doi.org/10.1103/PhysRevE.104.014145} {\bibfield
  {journal} {\bibinfo  {journal} {Phys. Rev. E}\ }\textbf {\bibinfo {volume}
  {104}},\ \bibinfo {pages} {014145} (\bibinfo {year} {2021})}\BibitemShut
  {NoStop}%
\bibitem [{\citenamefont {Giuliani}\ \emph
  {et~al.}(2017{\natexlab{b}})\citenamefont {Giuliani}, \citenamefont
  {Mastropietro},\ and\ \citenamefont {Toninelli}}]{Giuliani2017b}%
  \BibitemOpen
  \bibfield  {author} {\bibinfo {author} {\bibfnamefont {A.}~\bibnamefont
  {Giuliani}}, \bibinfo {author} {\bibfnamefont {V.}~\bibnamefont
  {Mastropietro}},\ and\ \bibinfo {author} {\bibfnamefont {F.~L.}\ \bibnamefont
  {Toninelli}},\ }\bibfield  {title} {\bibinfo {title} {Haldane relation for
  interacting dimers},\ }\href {https://doi.org/10.1088/1742-5468/aa5d1f}
  {\bibfield  {journal} {\bibinfo  {journal} {Journal of Statistical Mechanics:
  Theory and Experiment}\ }\textbf {\bibinfo {volume} {2017}},\ \bibinfo
  {pages} {034002} (\bibinfo {year} {2017}{\natexlab{b}})}\BibitemShut
  {NoStop}%
\bibitem [{\citenamefont {von Delft}\ and\ \citenamefont
  {Schoeller}(1998)}]{vonDelft1998}%
  \BibitemOpen
  \bibfield  {author} {\bibinfo {author} {\bibfnamefont {J.}~\bibnamefont {von
  Delft}}\ and\ \bibinfo {author} {\bibfnamefont {H.}~\bibnamefont
  {Schoeller}},\ }\bibfield  {title} {\bibinfo {title} {Bosonization for
  beginners — refermionization for experts},\ }\href@noop {} {\bibfield
  {journal} {\bibinfo  {journal} {Annalen der Physik}\ }\textbf {\bibinfo
  {volume} {7}},\ \bibinfo {pages} {225} (\bibinfo {year} {1998})}\BibitemShut
  {NoStop}%
\bibitem [{\citenamefont {Giuliani}\ \emph {et~al.}(2020)\citenamefont
  {Giuliani}, \citenamefont {Mastropietro},\ and\ \citenamefont
  {Toninelli}}]{Giuliani2020}%
  \BibitemOpen
  \bibfield  {author} {\bibinfo {author} {\bibfnamefont {A.}~\bibnamefont
  {Giuliani}}, \bibinfo {author} {\bibfnamefont {V.}~\bibnamefont
  {Mastropietro}},\ and\ \bibinfo {author} {\bibfnamefont {F.}~\bibnamefont
  {Toninelli}},\ }\bibfield  {title} {\bibinfo {title} {Non-integrable dimers:
  Universal fluctuations of tilted height profiles},\ }\href
  {https://doi.org/10.1007/s00220-020-03760-x} {\bibfield  {journal} {\bibinfo
  {journal} {Communications in Mathematical Physics}\ }\textbf {\bibinfo
  {volume} {377}} (\bibinfo {year} {2020})}\BibitemShut {NoStop}%
\bibitem [{\citenamefont {Powell}(2013)}]{Powell2013}%
  \BibitemOpen
  \bibfield  {author} {\bibinfo {author} {\bibfnamefont {S.}~\bibnamefont
  {Powell}},\ }\bibfield  {title} {\bibinfo {title} {Confinement of monopoles
  and scaling theory near unconventional critical points},\ }\href
  {https://doi.org/10.1103/PhysRevB.87.064414} {\bibfield  {journal} {\bibinfo
  {journal} {Phys. Rev. B}\ }\textbf {\bibinfo {volume} {87}},\ \bibinfo
  {pages} {064414} (\bibinfo {year} {2013})}\BibitemShut {NoStop}%
\bibitem [{\citenamefont {Sachdev}(2011)}]{Sachdev2011}%
  \BibitemOpen
  \bibfield  {author} {\bibinfo {author} {\bibfnamefont {S.}~\bibnamefont
  {Sachdev}},\ }\href@noop {} {\emph {\bibinfo {title} {Quantum phase
  transitions}}}\ (\bibinfo  {publisher} {Cambridge University Press},\
  \bibinfo {year} {2011})\BibitemShut {NoStop}%
\bibitem [{\citenamefont {Wilkins}(2021)}]{WilkinsThesis}%
  \BibitemOpen
  \bibfield  {author} {\bibinfo {author} {\bibfnamefont {N.}~\bibnamefont
  {Wilkins}},\ }\emph {\bibinfo {title} {Synchronization transition in the
  double dimer model}},\ \href {http://eprints.nottingham.ac.uk/65680/} {Ph.D.
  thesis},\ \bibinfo  {school} {University of Nottingham} (\bibinfo {year}
  {2021})\BibitemShut {NoStop}%
\bibitem [{\citenamefont {Alet}(2016)}]{Alet2016}%
  \BibitemOpen
  \bibfield  {author} {\bibinfo {author} {\bibfnamefont {F.}~\bibnamefont
  {Alet}},\ }\emph {\bibinfo {title} {{Dim{\`e}res classiques en interaction \&
  Autres probl{\`e}mes en magn{\'e}tisme quantique}}},\ \href
  {https://tel.archives-ouvertes.fr/tel-01798214} {\bibinfo {type}
  {Habilitation {\`a} diriger des recherches}},\ \bibinfo  {school}
  {{Universit{\'e} Paul Sabatier (Toulouse 3)}} (\bibinfo {year}
  {2016})\BibitemShut {NoStop}%
\bibitem [{\citenamefont {Nienhuis}(1987)}]{Nienhuis1987}%
  \BibitemOpen
  \bibfield  {author} {\bibinfo {author} {\bibfnamefont {B.}~\bibnamefont
  {Nienhuis}},\ }\bibfield  {title} {\bibinfo {title} {{C}oulomb gas
  formulation of two-dimensional phase transitions},\ }in\ \href@noop {} {\emph
  {\bibinfo {booktitle} {Phase transitions and critical phenomena}}},\
  Vol.~\bibinfo {volume} {11},\ \bibinfo {editor} {edited by\ \bibinfo {editor}
  {\bibfnamefont {C.}~\bibnamefont {Domb}}\ and\ \bibinfo {editor}
  {\bibfnamefont {J.}~\bibnamefont {Lebowitz}}}\ (\bibinfo  {publisher}
  {Academic},\ \bibinfo {address} {London},\ \bibinfo {year} {1987})\
  Chap.~\bibinfo {chapter} {1}\BibitemShut {NoStop}%
\bibitem [{\citenamefont {Allegra}(2015)}]{Allegra2015}%
  \BibitemOpen
  \bibfield  {author} {\bibinfo {author} {\bibfnamefont {N.}~\bibnamefont
  {Allegra}},\ }\bibfield  {title} {\bibinfo {title} {Exact solution of the 2d
  dimer model: Corner free energy, correlation functions and combinatorics},\
  }\href@noop {} {\bibfield  {journal} {\bibinfo  {journal} {Nuclear Physics
  B}\ }\textbf {\bibinfo {volume} {894}},\ \bibinfo {pages} {685} (\bibinfo
  {year} {2015})}\BibitemShut {NoStop}%
\bibitem [{\citenamefont {Hartwig}(1966)}]{Hartwig1966}%
  \BibitemOpen
  \bibfield  {author} {\bibinfo {author} {\bibfnamefont {R.~E.}\ \bibnamefont
  {Hartwig}},\ }\bibfield  {title} {\bibinfo {title} {Monomer pair
  correlations},\ }\href {https://doi.org/10.1063/1.1704931} {\bibfield
  {journal} {\bibinfo  {journal} {Journal of Mathematical Physics}\ }\textbf
  {\bibinfo {volume} {7}},\ \bibinfo {pages} {286} (\bibinfo {year}
  {1966})}\BibitemShut {NoStop}%
\bibitem [{\citenamefont {Chen}\ \emph {et~al.}(2009)\citenamefont {Chen},
  \citenamefont {Gukelberger}, \citenamefont {Trebst}, \citenamefont {Alet},\
  and\ \citenamefont {Balents}}]{Chen2009}%
  \BibitemOpen
  \bibfield  {author} {\bibinfo {author} {\bibfnamefont {G.}~\bibnamefont
  {Chen}}, \bibinfo {author} {\bibfnamefont {J.}~\bibnamefont {Gukelberger}},
  \bibinfo {author} {\bibfnamefont {S.}~\bibnamefont {Trebst}}, \bibinfo
  {author} {\bibfnamefont {F.}~\bibnamefont {Alet}},\ and\ \bibinfo {author}
  {\bibfnamefont {L.}~\bibnamefont {Balents}},\ }\bibfield  {title} {\bibinfo
  {title} {{C}oulomb gas transitions in three-dimensional classical dimer
  models},\ }\href {https://doi.org/10.1103/PhysRevB.80.045112} {\bibfield
  {journal} {\bibinfo  {journal} {Phys. Rev. B}\ }\textbf {\bibinfo {volume}
  {80}},\ \bibinfo {pages} {045112} (\bibinfo {year} {2009})}\BibitemShut
  {NoStop}%
\bibitem [{\citenamefont {Moodie}\ and\ \citenamefont
  {Long}(2020)}]{Moodie2020}%
  \BibitemOpen
  \bibfield  {author} {\bibinfo {author} {\bibfnamefont {J.~C.}\ \bibnamefont
  {Moodie}}\ and\ \bibinfo {author} {\bibfnamefont {M.~W.}\ \bibnamefont
  {Long}},\ }\bibfield  {title} {\bibinfo {title} {An exact power series
  representation of the {B}aker{\textendash}{C}ampbell{\textendash}{H}ausdorff
  formula},\ }\href {https://doi.org/10.1088/1751-8121/abcbae} {\bibfield
  {journal} {\bibinfo  {journal} {Journal of Physics A: Mathematical and
  Theoretical}\ }\textbf {\bibinfo {volume} {54}},\ \bibinfo {pages} {015208}
  (\bibinfo {year} {2020})}\BibitemShut {NoStop}%
\bibitem [{\citenamefont {Giamarchi}(2004)}]{Giamarchi2004}%
  \BibitemOpen
  \bibfield  {author} {\bibinfo {author} {\bibfnamefont {T.}~\bibnamefont
  {Giamarchi}},\ }\href@noop {} {\emph {\bibinfo {title} {Quantum physics in
  one dimension}}}\ (\bibinfo  {publisher} {Clarendon Press},\ \bibinfo {year}
  {2004})\BibitemShut {NoStop}%
\bibitem [{\citenamefont {Raghavan}\ \emph {et~al.}(1997)\citenamefont
  {Raghavan}, \citenamefont {Henley},\ and\ \citenamefont
  {Arouh}}]{Raghavan1997}%
  \BibitemOpen
  \bibfield  {author} {\bibinfo {author} {\bibfnamefont {R.}~\bibnamefont
  {Raghavan}}, \bibinfo {author} {\bibfnamefont {C.~L.}\ \bibnamefont
  {Henley}},\ and\ \bibinfo {author} {\bibfnamefont {S.~L.}\ \bibnamefont
  {Arouh}},\ }\bibfield  {title} {\bibinfo {title} {New two-color dimer models
  with critical ground states},\ }\href {https://doi.org/10.1007/bf02199112}
  {\bibfield  {journal} {\bibinfo  {journal} {Journal of Statistical Physics}\
  }\textbf {\bibinfo {volume} {86}},\ \bibinfo {pages} {517} (\bibinfo {year}
  {1997})}\BibitemShut {NoStop}%
\bibitem [{\citenamefont {Teber}(2007)}]{Teber2007}%
  \BibitemOpen
  \bibfield  {author} {\bibinfo {author} {\bibfnamefont {S.}~\bibnamefont
  {Teber}},\ }\bibfield  {title} {\bibinfo {title} {Bosonization approach to
  charge and spin dynamics of one-dimensional spin-$\frac{1}{2}$ fermions with
  band curvature in a clean quantum wire},\ }\href
  {https://doi.org/10.1103/PhysRevB.76.045309} {\bibfield  {journal} {\bibinfo
  {journal} {Phys. Rev. B}\ }\textbf {\bibinfo {volume} {76}},\ \bibinfo
  {pages} {045309} (\bibinfo {year} {2007})}\BibitemShut {NoStop}%
\bibitem [{\citenamefont {Pereira}\ \emph {et~al.}(2007)\citenamefont
  {Pereira}, \citenamefont {Sirker}, \citenamefont {Caux}, \citenamefont
  {Hagemans}, \citenamefont {Maillet}, \citenamefont {White},\ and\
  \citenamefont {Affleck}}]{Pereira2007}%
  \BibitemOpen
  \bibfield  {author} {\bibinfo {author} {\bibfnamefont {R.~G.}\ \bibnamefont
  {Pereira}}, \bibinfo {author} {\bibfnamefont {J.}~\bibnamefont {Sirker}},
  \bibinfo {author} {\bibfnamefont {J.-S.}\ \bibnamefont {Caux}}, \bibinfo
  {author} {\bibfnamefont {R.}~\bibnamefont {Hagemans}}, \bibinfo {author}
  {\bibfnamefont {J.~M.}\ \bibnamefont {Maillet}}, \bibinfo {author}
  {\bibfnamefont {S.~R.}\ \bibnamefont {White}},\ and\ \bibinfo {author}
  {\bibfnamefont {I.}~\bibnamefont {Affleck}},\ }\bibfield  {title} {\bibinfo
  {title} {Dynamical structure factor at small $q$ for the {XXZ} spin-1/2
  chain},\ }\href {https://doi.org/10.1088/1742-5468/2007/08/p08022} {\bibfield
   {journal} {\bibinfo  {journal} {Journal of Statistical Mechanics: Theory and
  Experiment}\ }\textbf {\bibinfo {volume} {2007}},\ \bibinfo {pages} {P08022}
  (\bibinfo {year} {2007})}\BibitemShut {NoStop}%
\bibitem [{\citenamefont {Fendley}\ \emph {et~al.}(2002)\citenamefont
  {Fendley}, \citenamefont {Moessner},\ and\ \citenamefont
  {Sondhi}}]{Fendley2002}%
  \BibitemOpen
  \bibfield  {author} {\bibinfo {author} {\bibfnamefont {P.}~\bibnamefont
  {Fendley}}, \bibinfo {author} {\bibfnamefont {R.}~\bibnamefont {Moessner}},\
  and\ \bibinfo {author} {\bibfnamefont {S.~L.}\ \bibnamefont {Sondhi}},\
  }\bibfield  {title} {\bibinfo {title} {Classical dimers on the triangular
  lattice},\ }\href {https://doi.org/10.1103/PhysRevB.66.214513} {\bibfield
  {journal} {\bibinfo  {journal} {Phys. Rev. B}\ }\textbf {\bibinfo {volume}
  {66}},\ \bibinfo {pages} {214513} (\bibinfo {year} {2002})}\BibitemShut
  {NoStop}%
\bibitem [{\citenamefont {Trousselet}\ \emph {et~al.}(2007)\citenamefont
  {Trousselet}, \citenamefont {Pujol}, \citenamefont {Alet},\ and\
  \citenamefont {Poilblanc}}]{Trousselet2007}%
  \BibitemOpen
  \bibfield  {author} {\bibinfo {author} {\bibfnamefont {F.}~\bibnamefont
  {Trousselet}}, \bibinfo {author} {\bibfnamefont {P.}~\bibnamefont {Pujol}},
  \bibinfo {author} {\bibfnamefont {F.}~\bibnamefont {Alet}},\ and\ \bibinfo
  {author} {\bibfnamefont {D.}~\bibnamefont {Poilblanc}},\ }\bibfield  {title}
  {\bibinfo {title} {Criticality of a classical dimer model on the triangular
  lattice},\ }\href {https://doi.org/10.1103/PhysRevE.76.041125} {\bibfield
  {journal} {\bibinfo  {journal} {Phys. Rev. E}\ }\textbf {\bibinfo {volume}
  {76}},\ \bibinfo {pages} {041125} (\bibinfo {year} {2007})}\BibitemShut
  {NoStop}%
\bibitem [{\citenamefont {Rossmann}(2006)}]{Rossmann2006}%
  \BibitemOpen
  \bibfield  {author} {\bibinfo {author} {\bibfnamefont {W.}~\bibnamefont
  {Rossmann}},\ }\href@noop {} {\emph {\bibinfo {title} {Lie Groups: An
  Introduction Through Linear Groups}}}\ (\bibinfo  {publisher} {Oxford
  University Press},\ \bibinfo {year} {2006})\BibitemShut {NoStop}%
\bibitem [{\citenamefont {Rao}\ and\ \citenamefont {Sen}(2001)}]{Rao2001}%
  \BibitemOpen
  \bibfield  {author} {\bibinfo {author} {\bibfnamefont {S.}~\bibnamefont
  {Rao}}\ and\ \bibinfo {author} {\bibfnamefont {D.}~\bibnamefont {Sen}},\
  }\bibinfo {title} {An introduction to bosonization and some of its
  applications},\ in\ \href {https://doi.org/10.1007/978-93-86279-07-1_6}
  {\emph {\bibinfo {booktitle} {Field Theories in Condensed Matter Physics}}}\
  (\bibinfo  {publisher} {Hindustan Book Agency},\ \bibinfo {address}
  {Gurgaon},\ \bibinfo {year} {2001})\ pp.\ \bibinfo {pages}
  {239--333}\BibitemShut {NoStop}%
\bibitem [{\citenamefont {Fetter}\ and\ \citenamefont
  {Walecka}(2003)}]{Fetter2003}%
  \BibitemOpen
  \bibfield  {author} {\bibinfo {author} {\bibfnamefont {A.~L.}\ \bibnamefont
  {Fetter}}\ and\ \bibinfo {author} {\bibfnamefont {J.~D.}\ \bibnamefont
  {Walecka}},\ }\href@noop {} {\emph {\bibinfo {title} {Quantum theory of
  many-particle systems}}}\ (\bibinfo  {publisher} {Dover Publications},\
  \bibinfo {address} {Mineola, N.Y},\ \bibinfo {year} {2003})\BibitemShut
  {NoStop}%
\bibitem [{\citenamefont {Rozhkov}(2005)}]{Rozhkov2005}%
  \BibitemOpen
  \bibfield  {author} {\bibinfo {author} {\bibfnamefont {A.~V.}\ \bibnamefont
  {Rozhkov}},\ }\bibfield  {title} {\bibinfo {title} {Fermionic quasiparticle
  representation of {T}omonaga-{L}uttinger {H}amiltonian},\ }\href@noop {}
  {\bibfield  {journal} {\bibinfo  {journal} {European Physical Journal B}\
  }\textbf {\bibinfo {volume} {47}},\ \bibinfo {pages} {193} (\bibinfo {year}
  {2005})}\BibitemShut {NoStop}%
\bibitem [{\citenamefont {Rozhkov}(2006)}]{Rozhkov2006}%
  \BibitemOpen
  \bibfield  {author} {\bibinfo {author} {\bibfnamefont {A.~V.}\ \bibnamefont
  {Rozhkov}},\ }\bibfield  {title} {\bibinfo {title} {Class of exactly soluble
  models of one-dimensional spinless fermions and its application to the
  {T}omonaga-{L}uttinger {H}amiltonian with nonlinear dispersion},\ }\href
  {https://doi.org/10.1103/PhysRevB.74.245123} {\bibfield  {journal} {\bibinfo
  {journal} {Phys. Rev. B}\ }\textbf {\bibinfo {volume} {74}},\ \bibinfo
  {pages} {245123} (\bibinfo {year} {2006})}\BibitemShut {NoStop}%
\bibitem [{\citenamefont {S{\'e}n{\'e}chal}(2004)}]{Senechal2004}%
  \BibitemOpen
  \bibfield  {author} {\bibinfo {author} {\bibfnamefont {D.}~\bibnamefont
  {S{\'e}n{\'e}chal}},\ }\bibinfo {title} {An introduction to bosonization},\
  in\ \href@noop {} {\emph {\bibinfo {booktitle} {Theoretical Methods for
  Strongly Correlated Electrons}}},\ \bibinfo {editor} {edited by\ \bibinfo
  {editor} {\bibfnamefont {D.}~\bibnamefont {S{\'e}n{\'e}chal}}, \bibinfo
  {editor} {\bibfnamefont {A.-M.}\ \bibnamefont {Tremblay}},\ and\ \bibinfo
  {editor} {\bibfnamefont {C.}~\bibnamefont {Bourbonnais}}}\ (\bibinfo
  {publisher} {Springer New York},\ \bibinfo {year} {2004})\ pp.\ \bibinfo
  {pages} {139--186}\BibitemShut {NoStop}%
\end{thebibliography}%

\end{document}